%% file: ic.tex
\documentclass[usenatbib]{mn2e}
\usepackage{amsfonts}
\usepackage{amssymb}
\usepackage{amsmath}
\usepackage{graphicx}
\usepackage{fixltx2e} %%forces correct figure order-breaks \thanks w/ pdflatex
\usepackage{bm}

\input{defsDarren.tex}

  \title[Toward an accurate mass function]
  {Toward an accurate mass function for precision cosmology}

\author[Reed \etal]{\parbox{\textwidth}{
Darren S. Reed$^{1*}$, %% \thanks is broken by \usepackage{fixltx2e}
Robert E. Smith$^{1,2}$, 
Doug Potter$^{1}$,
Aurel Schneider$^{1}$\\
Joachim Stadel$^{1}$
\&
Ben Moore$^{1}$}\vspace{0.4cm}\\
\parbox{\textwidth}{ 
$^1$Insitute for Theoretical Physics, Univ. of
  Z\"{u}rich, Winterthurerstrasse 190, CH-8057 Z\"{u}rich, Switzerland\\
  $^2$Argelander-Institute for Astronomy, Auf dem H\"ugel 71, D-53121
  Bonn, Germany }}

\begin{document}

\pagerange{\pageref{firstpage}--\pageref{lastpage}}

\pubyear{2012}

\maketitle

\label{firstpage}

%%%%%%%%%%%%%%%%%%%%%%%%%%%%%%%%%%%%%%%%%%%%%%%%%%%%%%%

\begin{abstract}
  Cosmological surveys aim to use the evolution of the abundance of
  galaxy clusters to accurately constrain the cosmological model.  In
  the context of $\Lambda$CDM, we show that it is possible to achieve
  the required percent level accuracy in the halo mass function with
  gravity-only cosmological simulations, and we provide simulation
  start and run parameter guidelines for doing so.  Some previous
  works have had sufficient statistical precision, but lacked robust
  verification of absolute accuracy.  Convergence tests of the mass
  function with, for example, simulation start redshift can exhibit
  false convergence of the mass function due to counteracting errors,
  potentially misleading one to infer overly optimistic estimations of
  simulation accuracy.  Percent level accuracy is possible if initial
  condition particle mapping uses second order Lagrangian Perturbation
  Theory, and if the start epoch is between 10 and 50 expansion
  factors before the epoch of halo formation of interest.  The mass
  function for halos with fewer than $\sim 1000$ particles is highly
  sensitive to simulation parameters and start redshift, implying a
  practical minimum mass resolution limit due to mass discreteness.
  The narrow range in converged start redshift suggests that it is not
 presently possible for a single simulation to capture accurately the cluster
  mass function while also starting early enough to model accurately
  the numbers of reionisation era galaxies, whose baryon feedback
  processes may affect later cluster properties.  Ultimately, to fully
  exploit current and future cosmological surveys will require
  accurate modeling of baryon physics and observable properties, a
  formidable challenge for which accurate gravity-only simulations are
  just an initial step.

\end{abstract}

%%%%%%%%%%%%%%%%%%%%%%%%%%%%%%%%%%%%%%%%%%%%%%%%%%%%%%%


%%%%%%%%%%%%%%%%%%%%%%%%%%%%%%%%%%%%%%%%%%%%%%%%%%%%%%%
%%%%%%%%%%%%%%%%%%%%%%%%%%%%%%%%%%%%%%%%%%%%%%%%%%%%%%%

\section{introduction}

%% this replaces the broken \thanks
\let\thefootnote\relax\footnotetext{$^*$email: \texttt{ reed@physik.uzh.ch}}

In the vacuum energy dominated cold dark matter cosmological model
\citep[hereafter $\Lambda$CDM,][]{Komatsuetal2011short}, large-scale
structures form through the amplification of small density
fluctuations via gravitational instability. At early times this
amplification can be followed using linear perturbation theory of the
general relativistic equations of motion for the field. At late times,
owing to the nonlinearities in the equations, and after
shell-crossing, the dynamics may only be accurately followed using
numerical simulations.  Overdense regions of the density field, whose
dynamics have broken away from the evolution of the background
space-time and have reached some state of virial equilibrium are
commonly referred to as dark matter halos.

The growth rate of large-scale structures is directly sensitive to the
expansion rate of the Universe, and hence the cosmological
parameters. One can show theoretically, through the excursion set
formalism \citep{PressSchechter1974,Bondetal1991,ShethTormen1999},
that the number of halos is also sensitive to cosmological parameters,
and importantly for future surveys, the presence of ``dark energy''
\citep{WangSteinhardt1998,Haimanetal2001,LimaHu2004,MarianBernstein2006,Cunhaetal2010,Courtinetal2011}. This
forecast cosmological sensitivity has recently been verified through
direct testing with $N$-body simulations \citep{SmithMarian2011}.

The amount of cosmological information that can be extracted from
cluster number counts is limited by our ability to detect
signal-to-noise peaks in our observational survey -- \ie associate
galaxies to groups, identify groups relative to an X-ray background
noise level, etc. The lower that one can push the minimum detectable
mass, the more cosmological information can be extracted from the
survey. This comes under the proviso that one can accurately calibrate
the true--observable mass relationship
\citep{LimaHu2005,Marianetal2009,Rozoetal2009,Mandelbaumetal2010,OguriTakada2011,Anguloetal2012}. The
numbers of rare halos are also sensitive to the level of
non-Gaussianity in the primordial density field due to its effect upon
the tail of extreme density fluctuations
\citep{Matarreseetal2000,Marianetal2011a}. 
Cluster counts are also sensitive to the total neutrino mass \citep[\eg]{Wangetal2005,Carboneetal2012,Shimonetal2012}
Thus, surveys that promise
to accurately measure the evolution of the abundance of groups and
clusters, also have the potential to help probe fundamental physics.
Accurate theoretical predictions for the cluster mass function and its
dependence on cosmology, are therefore essential to fully exploit next
generation cluster surveys.

Current cosmological constraints from clusters come from:
\citet{Vikhlininetal2009,Vanderlindeetal2010short,Rozoetal2010,
Sehgaletal2011short,Allenetal2011,Planck2011SZshort}.  Over the next
decade there will be a number of large surveys that will aim to
strongly constrain the cosmological model through the abundance of
clusters: in the X-ray there will be eROSITA
\citep{Pillepichetal2012}, with the Sunyaev-Zel'Dovich method there
will be Planck\footnote{www.rssd.esa.int/planck}, in the optical using
the weak lensing method there will be
DES\footnote{www.darkenergysurvey.org}, Euclid \citep{EUCLID2011} and
LSST\footnote{www.lsst.org/lsst}.  Several authors have estimated the
requirements on the theoretical accuracy of the halo mass function to
achieve the statistically limited constraints on cosmological
parameters.  \citet{Wuetal2010} point out that, in order to constrain
time evolving dark energy models for DES, the theoretical mass
function must be known with an accuracy $\lesssim0.5\%$.

In this paper we address the question: What are the correct numerical
parameters needed to achieve percent level accuracy in the mass
function in a cosmological simulation?  Large simulation volumes
(whether by a single simulation or by multiple realizations) are able
to reduce statistical uncertainties due to finite halo
numbers. However, large absolute volumes are needed to reduce
systematic and statistical errors associated with poor sampling of
large-scale density modes
\citep[e.g.][]{BarkanaLoeb2004,BaglaRay2005,PowerKnebe2006,Reedetal2007b,Crocceetal2010,SmithMarian2011}.
Over the past decade, impressive statistical precision in the halo
mass function has been achieved using suites of cosmological
simulations \citep[][and
others]{Jenkinsetal2001,Warrenetal2006,Reedetal2007b,Tinkeretal2008,Crocceetal2010,Ilievetal2012,Bhattacharyaetal2011,SmithMarian2011,Anguloetal2012,Alimietal2012,Watsonetal2012}.
However, statistical precision does not imply accuracy, even when
considering gravity-only simulations.  Sources of systematic error
include finite simulation volume, force resolution, mass resolution
and discreteness effects, time-stepping, halo finding, initial
condition particle mapping, and start redshift.  Recent progress
includes \citet{Crocceetal2010} and \citet{Bhattacharyaetal2011}, who
each address many of the systematic uncertainties and determine a halo
mass function with an estimated accuracy of $\sim2\%$ from a suite of
large gravity-only cosmological boxes, though their results differ by
significantly more than this for halos larger than $\sim
10^{15} \Msol$. Moreover, neither approach has taken into account the
full covariance matrix of mass function estimates when deriving best
fit parameters \citep{SmithMarian2011}.

As a first step on the path toward producing an accurate mass
function in the observational plane, we limit ourselves to
demonstrating how percent level accuracy in gravity-only (\ie
collisionless) simulations (wherein baryons are present but interact
only via gravity) can be accomplished.  We show how to set up initial
conditions so that percent level accuracy can be achieved.  We also
isolate and test the run parameters that control force resolution
(force softening and tree opening angle) and time-step size, allowing
us to determine the required values to achieve percent level
convergence.  Finally, we ask: how many particles do we need to sample
the halo mass distribution, in order to obtain a mass function
accurate to better than $\lesssim1\%$.

The paper breaks down as follows: in \S\ref{sec-techniques} we discuss
setting up the initial conditions for the structure
formation simulations and the parameters used to run the $N$-body
codes.  In \S\ref{sec-simulations} we describe the suite of $N$-body
simulations performed and halo identification. In \S\ref{sec-mf} we
present the results for the halo mass function and its convergence
with simulation parameters. In \S\ref{sec-ic} we explore the variation
of the halo mass function with the method for generating the initial
conditions. We also make a comparison between the results obtained
from two well known $N$-body codes. In \S\ref{sec-conother} we explore
the convergence of the matter power spectrum and the 1-point
probability density function of matter fluctuations. In
\S\ref{sec-challenges} we discuss the remaining challenges of
obtaining better than 1\% accurate mass functions from structure
formation simulations. In \S\ref{sec-guidelines} we summarize our
findings and draw up a set of guidelines for obtaining accurate 
gravity-only mass functions.

%%%%%%%%%%%%%%%%%%%%%%%%%%%%%%%%%%%%%%%%%%%%%%%%%%%%%%%
%%%%%%%%%%%%%%%%%%%%%%%%%%%%%%%%%%%%%%%%%%%%%%%%%%%%%%%

\section{Simulating structure formation}
\label{sec-techniques}

%%%%%%%%%%%%%%%%%%%%%%%%%%%%%%%%%%%%%%%%%%%%%%%%%%%%%%%

\subsection{initial conditions}
\label{sec-icsetup}

In order to set up a simulation, we must first select the cosmological
model and the probability distribution of the primordial density
perturbations. In this study we shall work within the context of the
$\Lambda$CDM paradigm and assume that the initial density modes are
described by a Gaussian random field. The statistics of the field are
thus fully specified by the power spectrum. Hence, a particular
realization of the density field in Fourier space may be obtained by
drawing a set of uniform random phases and assigning amplitudes drawn
from the Rayleigh distribution \citep{Efstathiouetal1985}, or through
the convolution of white noise with a filter that is related to the
power spectrum \citep{Bertschinger2001}.

A given density field must then be converted into a particle
distribution, and several techniques for doing this have been discussed
in the literature
\citep[\eg][]{Efstathiouetal1985,Scoccimarro1998,Bertschinger1999,Bertschinger2001,Crocceetal2006}.
The traditional approach is to place particles on a uniform lattice,
and these are then displaced off the initial points $\bq$ using a
displacement field ${\bm\Psi}(\bq)$ that encodes all of the
statistical properties of the density field. Hence, initial
(Lagrangian) and final (Eulerian) positions, {\bf x}, are related
through:
\be {\bf x} = {\bf q} + {\bm\Psi}(\bq,\tau)\ ,\ \ee
where the coordinates $\bx$ are a solution to the equation of motion:
\be 
\frac{d^2{\bx}}{d\tau^2}+{\mathcal H}(\tau)\frac{d\bx}{d\tau} = -\nabla \Phi 
\label{eq:eom}\ee
where in the above $d\tau=dt/a(t)$ is conformal time, ${\mathcal
  H}=aH(a)$, and $\Phi$ is the peculiar gravitational potential.  The
solution for $\Psi$ is perturbative, and each order can be found
through iteration with solutions of lower order. In terms of the
initial density field, and up to second order, the solutions may be
written
\citep{Zeldovich1970,Buchert1994,Buchertetal1994,Bouchetetal1995,Scoccimarro1998}:
\be 
{\bm\Psi}(\bq,\tau) = - D_1(a){\bm\nabla}_q\phi^{(1)}(\bq) + D_2(a){\bm\nabla}_q\phi^{(2)}(\bq), 
\label{eqn-2lpt}
\ee
where $D_1(a)$ and $D_2(a)\approx -3D_1^2(a)/7$ are the first and
second order growth factors suitable for $\Lambda$CDM.  The potentials
$\phi^{(1)}(\bq)$ and $\phi^{(2)}(\bq)$ can be found through
iteratively solving the Poisson equations:
\ba
\nabla^2_q\phi^{(1)}({\bf q}) & = &  \delta^{(1)}({\bf q}) \ ;\\
\nabla^2_q\phi^{(2)}({\bf q}) & = &  \sum_{i>j}^{3} 
\left\{\phi_{,ii}^{(1)}(\bq)\phi_{,jj}^{(1)}(\bq)-\left[\phi^{(1)}_{,ij}\right]^2\right\} \ ,
\label{eqn-poisson}
\ea
where $\phi_{,ij}\equiv \partial^2\phi/\partial q_i\partial q_j$. The
linear solutions for $\bm\Psi$, with $\phi^{(2)}=0$, yield the
traditional Zel'Dovich approximation, which we refer to as 1LPT, and
the second order solutions, with $\phi^{(2)}$, we refer to as
2LPT. \citet{Scoccimarro1998} gave a detailed prescription for
implementing 2LPT displacements in simulations, and we make use of a
slightly modified version of the publicly available code {\tt
  2LPT}\footnote{http://cosmo.nyu.edu/roman/2LPT}.  \citet{Crocceetal2006}
demonstrated that 2LPT reduces numerical ``transients'' to the level where
an accurate representation of the halo mass function may be obtained
for relatively late start times, $a_{\rm f}/a_{\rm i}\approx 10$,
where $a_{\rm i}$ and $a_{\rm f}$ are the initial and final expansion
factors.

As can be seen from \Eqn{eqn-2lpt}, in the limit of asymptotically
high initial redshift 1LPT and 2LPT become equivalent since
$D_2(a_{\rm i})/D_2(a_{\rm f})\ll D_1(a_{\rm i})/D_1(a_{\rm f})$. This
has led some to speculate that, provided one takes the initial start
redshift to be sufficiently high, then it should not matter whether
one uses 1LPT or 2LPT.  This issue will be investigated in detail in
\S\ref{sec-ic}.

Several earlier studies have explored the importance of 1LPT versus 2LPT
initial conditions: \citet{Knebeetal2009} used {\tt Gadget-2} to show
that start redshift and 2LPT versus 1LPT had little effect on internal
halo properties, specifically testing halo concentration, spin
parameter, tri-axiality.  They also found little dependence on halo
mass or the halo mass function.  However, their results may understate
any numerical issues because they focused on smaller halos of
$10^{10}$--$10^{13}\Msol$ where the mass function is not very steep,
and their statistics were limited due to using low-resolution
$N=256^3$ particles. A more recent study by \citet{Jenkins2010}, has
shown that there is a definite, although weak, dependence of the
subhalo mass function inside Milky-Way mass halos on the choice of
the initial conditions.

One last issue, concerning the generation of initial Gaussian random
density fields, is that some researches have advocated the use of the
``Hann filter''\footnote{The ``Hann filter'' is sometimes
  (erroneously, according to Wikipedia) referred to as the ``Hanning''
  filter.} \citep{Bertschinger2001}. This is an anti-aliasing filter
\citep{Pressetal1992}, and corresponds to setting the Fourier density
modes that are outside the Nyquist sphere of the simulation, $k_{\rm
  Ny}=\pi N^{1/3}/L$, to vanish by multiplying the transfer function
by $W(k)=\cos(\pi k/2k_{\rm Ny})$ for $k<k_{\rm Ny}$ and 0 for
$k>k_{\rm Ny}$. The purpose of this is to mitigate some of the
anisotropies in the forces due to the cubical lattice. In
\S\ref{sec-ic} we shall also investigate how the presence of such
filtering impacts our goal of an accurate mass function.

Note that for some of the simulations where we test for parameter
convergence, we will also make use of a modified version of {\tt
  Grafic-2} \citep{Bertschinger2001}.  These two initial condition
codes were verified to show identical convergence trends with start
redshift.

%%%%%%%%%%%%%%%%%%%%%%%%%%%%%%%%%%%%%%%%%%%%%%%%%%%%%%%

\subsection{$N$-body codes}
\label{sec-haloprofile}

Once we have obtained an initial condition, we then need to integrate
the equations of motion, \Eqn{eq:eom}. In this study we shall make use
of two standard $N$-body techniques: {\tt PKDGRAV} V2.2.12 and {\tt
  Gadget-2}.

\vspace{0.1cm}
\noindent {\tt PKDGRAV} is our primary simulation code for this study,
an early version of which is described in \citet{Stadel2001}. The
version of the code we use has been MPI parallelized, and uses the
hierarchical tree data structure to organize the individual simulation
particles.  The gravitational force on each particle is calculated
using a multipole expansion with Ewald summation to replicate the
simulation cube as an approximation of an infinite periodic universe.
The peculiar potential around any given particle is obtained from an
hexadecapole expansion of the forces. {\tt PKDGRAV} uses a variable
time step criterion that is synchronized for global
time-steps. Particle orbits are integrated with the symplectic leapfrog
integrator.

\vspace{0.1cm}
\noindent {\tt Gadget-2} is a tree-particle-mesh (Tree-PM) code, and
full details of which may be found in \citet{Springel2005}. The main
difference with {\tt PKDGRAV} is that on large scales it uses Fourier
based methods to solve for the forces and only uses the tree algorithm
to solve for forces on small scales. The solution for the potential is
then obtained through an interpolation of the PM and tree forces over
the force matching region, and typically this is $\sim4-5$ mesh cells.

\vspace{0.1cm} In \S\ref{sec-ic} we investigate the mass functions
from these different codes and explore the convergence properties with
different 1LPT and 2LPT start redshifts.  Additionally, we aim to
determine the typical values for ``generic'' run parameters that are
required for percent level convergence.  In what follows we shall
describe parameters that are mainly specific to {\tt PKDGRAV}, but
will make reference as to how they apply to {\tt Gadget-2} or other codes. 
{\tt Gadget-2} parameters are tested further in \citet{Smithetal2012}. 
The run parameters that we focus on are:

\vspace{0.1cm}
\noindent {\bf Tree opening angle $\Theta$}: The tree opening angle
controls the accuracy of medium and long range forces. It does this by
setting the minimum distance between a given particle and tree node
below which the tree node will be ``opened''. Thus the force
calculations for a given particle will include contributions from
entire nodes and or individual particles.  A discussion of how
$\Theta$ relates to other tree types can be found in
\citet{Stadel2001}.

\vspace{0.1cm}
\noindent {\bf Softening $\epsilon$}: In order to avoid excessively
large accelerations and hence excessively short time-steps, the
small-scale gravitational interactions must be ``softened''. This
makes sense for simulations of collisonless systems like CDM, where
the particles represent large coarse grained elements of the
microscopic phase space. {\tt PKDGRAV} and {\tt Gadget-2} both use a
softened kernel: gravitational forces approach zero for spatially
coinciding particles, and become Newtonian at $2\epsilon$ for {\tt
  PKDGRAV} and $2.8\epsilon$ for {\tt Gadget-2}.  
{\tt PKDGRAV} uses the K$_3$ softening kernal of \citet{Dehnen2001} while {\tt Gadget-2} uses a spline kernel.
The force softening
leads to a minimum resolved spatial scale. 
Throughout, we make use of
constant comoving softening.

\vspace{0.1cm}
\noindent {\bf Time-step $\eta$}: Each particle is on an adaptive
time-step with length proportional to the time-step parameter $\eta$.
The actual time-step length for each particle is based on the magnitude
of its current acceleration $|{\bf a}|$, the softening length
$\epsilon$, and the time-step parameter $\eta$, in accordance with the
relation:
\be
{\rm dt \ge \eta \sqrt{(\epsilon/|{\bf a}|)}} \ .
\ee
This technique allows significant computational savings in
cosmological simulations when only a small fraction of particles are
in dense regions requiring the shortest time-steps.
 
\vspace{0.2cm} In summary, we shall investigate how the halo mass
function varies with: the initial start redshift; with 1LPT or 2LPT
initial conditions; with Nyquist filtering; we shall explore results
for two simulation codes; and how variations in $\Theta, \epsilon,
\eta$ affect our results. Besides these, we shall also explore finite
volume effects and mass resolution.

%%%%%%%%%%%%%%%%%%%%%%%%%%%%%%%%%%%%%%%%%%%%%%%%%%%%%%%

\begin{table*}
\begin{center}
\begin{tabular}{c|c|c|c|c|c|c|c|c|c|}
  Code & IC method & $L\, [\Mpc]$ & ${\rm m_{p}}\, [\Msol]$ & $z_{\rm i} $ & $z_{\rm f}$ & $\eta$ 
  & ${\rm \epsilon}$ & $\Theta$ \\
  \hline

\hline
{\tt PKDGRAV} & 1LPT & 17.625 & $3.88\times10^5$ & 30, 49, 100, 200, 400 &  9.8 & 0.15 & ${l_{\rm m}/50}$ & 0.55 & \\

\hline
{\tt PKDGRAV} & 2LPT & 17.625 & $3.88\times10^5$ & 30, 49, 100, 200, 400 &  9.8& 0.15 & ${\lm/50}$ & 0.55 & \\

\hline
{\tt Gadget-2} & 1LPT &  17.625  & $3.88\times10^5$ & 30, 49, 100, 200, 400 &  9.8& 0.2  & ${\lm/30}$ & 0.5 & \\
{\tt Gadget-2} & 2LPT & 17.625 & $3.88\times10^5$ & 30, 49, 100, 200, 400 &  9.8& 0.2  & ${\lm/30}$ & 0.5 & \\

\hline
{\tt PKDGRAV} & 2LPT & 17.625 & $3.88\times10^5$ & 400 &  9.8& 0.15 & ${\lm/50}$ & 0.4 & \\
{\tt PKDGRAV} & 2LPT & 17.625 & $3.88\times10^5$ & 400 &  9.8& 0.15 & ${\lm/50}$ & 0.68 & \\
{\tt PKDGRAV} & 2LPT & 17.625 & $3.88\times10^5$ & 400 &  9.8& 0.15 & ${\lm/50}$ & 0.8 & \\

\hline
{\tt PKDGRAV} & 2LPT & 17.625 & $3.88\times10^5$ & 400 &  9.8& 0.07 & ${\lm/50}$ & 0.55 & \\
{\tt PKDGRAV} & 2LPT & 17.625 & $3.88\times10^5$ & 400 &  9.8& 0.2 & ${\lm/50}$ & 0.55 & \\
{\tt PKDGRAV} & 2LPT & 17.625 & $3.88\times10^5$ & 400 &  9.8& 0.25 & ${\lm/50}$ & 0.55 & \\
{\tt PKDGRAV} & 2LPT & 17.625 & $3.88\times10^5$ & 400 &  9.8& 0.3 & ${\lm/50}$ & 0.55 & \\
{\tt PKDGRAV} & 2LPT & 17.625 & $3.88\times10^5$ & 400 &  9.8& 0.6 & ${\lm/50}$ & 0.55 & \\

\hline
{\tt PKDGRAV} & 1LPT-g & 17.625 & $3.88\times10^5$ & 400 &  9.8& 0.15 & ${\lm/50}$ & 0.55 & \\
{\tt PKDGRAV} & 1LPT-g-HF & 17.625 & $3.88\times10^5$ & 400 &  9.8& 0.15 & ${\lm/50}$ & 0.55 & \\
{\tt PKDGRAV} & 1LPT-g & 17.625 & $3.88\times10^5$ & 400 &  9.8& 0.15 & ${\lm/20}$ & 0.55 & \\
{\tt PKDGRAV} & 1LPT-g & 17.625 & $3.88\times10^5$ & 400 &  9.8& 0.15 & ${\lm/10}$ & 0.55 & \\
{\tt PKDGRAV} & 1LPT-g & 17.625 & $3.88\times10^5$ & 400 &  9.8& 0.15 & ${\lm/5}$ & 0.55 & \\
{\tt PKDGRAV} & 1LPT-g & 17.625 & $3.88\times10^5$ & 400 &  9.8& 0.15 & ${\lm/2}$ & 0.55 & \\

\hline
{\tt PKDGRAV} & 1LPT & 2048 & $6.05\times10^{11}$ & 30, 49, 200 &  0.0 & 0.15 & ${\lm/50}$ & 0.55($z>2$),0.7($z<2$)   &\\
{\tt PKDGRAV} & 2LPT & 2048 & $6.05\times10^{11}$ & 30, 49, 100, 200, 400 & 0.0 & 0.15 & ${\lm/50}$ & 0.55($z>2$),0.7($z<2$) & \\
{\tt PKDGRAV} & 1LPT-g & 2048 & $6.05\times10^{11}$ &  400 & 0.0 & 0.15 & ${\lm/50}$ & 0.55($z>2$),0.7($z<2$) & \\
{\tt Gadget-2} & 2LPT & 2048 & $6.05\times10^{11}$ & 30, 200 & 0.0 & 0.2 & ${\lm/20}$ & 0.5 & \\

\hline
\end{tabular} 
\caption{Suite of cosmological simulations.  Col.~1: $N$-body code
  used. Col.~2: initial condition method, note that we used the {\tt
    2LPT} code throughout, except where there is a {\it -g}, which
  denotes the use of {\tt Grafic-2}; {\it -HF} denotes use of a Hann
  filter.  Col.~3: box size. Col.~4: particle mass. Col.~5: initial
  condition start redshifts that have been simulated. Col.~6: final
  redshift. Col.~7: time-stepping parameter. Col.~8: Force softening,
  ${\rm \epsilon}$, in units of the mean inter-particle spacing
  $\lm$. Col.~9: tree opening angle ({\tt ErrTolTheta} for {\tt
    Gadget-2} runs).  The additional {\tt Gadget-2} parameters were
  set to: {\tt ErrTolIntAccuracy}=0.02, {\tt
    MaxRMSDisplacementFac}=0.2, {\tt MaxSizeTimestep}=0.02, {\tt
    MinSizeTimestep}=0.000, {\tt ErrTolForceAcc}=0.005, {\tt
    TreeDomainUpdateFrequency}=0.05, {\tt PMGRID=1024}. }
\label{table-sims}
\end{center}  
\end{table*}

%%%%%%%%%%%%%%%%%%%%%%%%%%%%%%%%%%%%%%%%%%%%%%%%%%%%%%%
%%%%%%%%%%%%%%%%%%%%%%%%%%%%%%%%%%%%%%%%%%%%%%%%%%%%%%%

\section{Simulations}
\label{sec-simulations}

\subsection{Simulation suite}

We have generated a suite of $N$-body simulations that are designed to
explore the accuracy with which we may estimate the halo mass function
and its dependence on how we simulate the dark matter (as discussed in
\S\ref{sec-techniques}).  All of the simulations that we have
performed evolve $N=1024^{3}$ equal mass dark matter particles. We
consider periodic cubes of size $L=17.625 \Mpc$ evolved to $z=10$, and
cubes of size $L=2048\Mpc$ evolved to $z=0$.  The relative box sizes
were chosen so that halos corresponding to $\sim3\sigma$ fluctuations in the
density field are sampled by $N_{\rm h}\sim 1000$ particles at the
final output. This corresponds to $M\sim 3.8\times 10^{8} \Msol$ for
the small boxes at $z=10$, and $M\sim 6.1\times10^{14} \Msol$ at
$z=0$ for the larger boxes. Thus, the final halos in the small box
simulations are in an evolutionary state similar to the clusters in
the large box simulations at $z=0$.

Although the $L=17.625 \Mpc$ box is very small, the effects of finite
volume on our study are attenuated because we examine halos early, at
$z=10$, when the typical halo mass-scale is still much smaller than
the total simulation mass.  There is no need to apply a finite volume
correction as in \eg \citet{Reedetal2007b} to these simulations
because each of our convergence series utilizes identical initial
conditions and particle displacements.  Finite volume effects, to
the extent that they can be accounted for with a simple linear
correction technique, are thus identical within each convergence
series.  

The cosmological parameters that we adopted for the small box runs
were consistent with WMAP5 \citep{Komatsuetal2009short}:
$\Omega_m=0.274$, $\Omega_{\Lambda}=0.726$, $\Omega_b=0.046$,
$h=0.705$, $n_s=0.96$, $\sigma_8=0.812$, where these parameters are
the density parameters in matter, vacuum energy, and baryons; the
dimensionless Hubble parameter; the primordial power spectral index;
and the variance of the density fluctuations on scales of $R=8\Mpc$.  The
transfer function that we used to create the initial conditions was
produced using the prescription of \citet{EisensteinHu1998}.  For the
large box runs, the cosmological parameters we adopted were consistent
with WMAP7 \citep{Komatsuetal2011short}: $\Omega_m=0.2726$;
$\Omega_{\Lambda}=0.7274$, $\Omega_b=0.046$, $h=0.704$, $n_s=0.963$,
$\sigma_8= 0.809$. The transfer function for these runs was computed
using {\small CAMB} \citep{Lewisetal1999}. Full details for the entire
suite of simulations are given in Table~\ref{table-sims}.

%%%%%%%%%%%%%%%%%%%%%%%%%%%%%%%%%%%%%%%%%%%%%%%%%%%%%%%

\subsection{Halo identification}
\label{sec-halofinding}

We identify dark matter halos using the {\it friends-of-friends} (FoF)
algorithm \citep{Davisetal1985}. This links together all particles
that are separated by less than the linking length $b$, where $b$ is
expressed in units of the mean inter-particle separation. Throughout,
we use $b=0.2$, and we use the particular implementation of the
algorithm internal to {\tt PKDGRAV}; a similar implementation is
provided by the code {\tt
  fof}\footnote{www-hpcc.astro.washington.edu/tools/fof.html}. The
estimated iso-density contour that the FoF recovers is
$\delta=\delta\rho/\rhob = n_{\rm c} b^{-3}-1 \approx 81.62$
\citep{Moreetal2011}.

In the literature there is a wide range of alternate approaches to the
identification of dark matter halos: \eg the {\it spherical
  over-density} (SO) algorithm \citep{LaceyCole1994,Tinkeretal2008};
6-D phase space algorithms \citep{Behroozietal2013}; and for a review
of methods see \citet{Knebeetal2011short}. Rather than explore all of
these different methods here, we shall work under the assumption that:
an accurate FoF mass function implies an accurate mass function for
all other algorithms designed to select {\em approximately} virialised
objects. Anecdotal support for this is provided by \citet{McBrideetal2011b}, who
found similar convergence behavior with simulation set-up for the FoF
($b=0.2$) and SO 200 mass functions.  We shall reserve the task of
establishing the veracity of this assumption for future work.

A number of systematic errors have been noted for halos obtained with
the FoF algorithm. Firstly, the recovered halo masses are
systematically overestimated with respect to the ``true'' halo mass
\citep{Warrenetal2006,Lukicetal2009,Trentietal2010,Moreetal2011}.
This owes to the fact that the true halo mass distribution is sampled
by a finite number of particles.  This mass overestimation has been
quantified for spherical mock halos by \citet{Lukicetal2009} and more
recently by \citet{Moreetal2011}. Secondly, FoF halos also experience
``bridging'', which may occur when two distinct halos undergo a close
encounter. This systematic effect: links unvirialised systems, it
acts to reduce the overall number of halos found, and it predominantly
occurs for the highest mass objects and is stronger at higher
redshifts \citep{Lukicetal2009}.

\citet{Warrenetal2006} and \citet{Bhattacharyaetal2011} have provided
empirical corrections, determined from cosmological simulations, for
the systematic FoF errors.  However, these empirical models are mass
and redshift independent, which may make them insufficient for our
goal of achieving a $\sim1\%$ accurate mass function. We would expect
the FoF errors to include dependencies on the specific distribution of
mass within halos, which depends on both mass and redshift.
Nevertheless, we assert that the FoF mass overestimation should not
affect our convergence tests because they are all performed at the
same mass resolution.  For these reasons, we use the raw FoF masses,
and remark that this will affect our ability to recover an
``unbiased'', FoF mass function. However, this is sufficient to our
purposes of quantifying relative differences in the mass function of
different simulations.

%%%%%%%%%%%%%%%%%%%%%%%%%%%%%%%%%%%%%%%%%%%%%%%%%%%%%%%
%%%%%%%%%%%%%%%%%%%%%%%%%%%%%%%%%%%%%%%%%%%%%%%%%%%%%%%

\section{Mass Function Convergence I: Simulation Parameters}
\label{sec-mf}

In this section, we show convergence of the FoF halo mass function
with varying simulation run and set-up parameters.  We estimate the
number of halos per logarithmic mass interval, $dn(M)/d{\rm
  log_{10}}M$, using a Gaussian kernel in log mass.  The Gaussian
kernel is convenient for diagnostics because it avoids the `saw-tooth'
pattern that emerges in a binned mass function as a result of the
discretization of halo masses, particularly at low masses where the
simulation particle mass is a significant fraction of the mass-width
of a bin.  The number of halos with mass $M_k$ is estimated by the
sum:
\be 
N_k = \frac{\sum_i M_{hi} f_g}{\sum_i M_{hi}} \ ,
\label{eqn-kernel}
\ee
where $f_g$ is a Gaussian kernel (in $log_{10}M$) of width $h=0.0625$, chosen
to minimize Poisson fluctuations without introducing systematic error
to the mass function.  To minimize computational cost, we truncate the
kernel beyond a range of $\pm a=3h$.  The number of halos per
logarithmic mass interval per unit volume is:
\be
\frac{dn(M)}{d{\rm log_{10}}M} = \frac{N_k}{V h ~{\rm erf}(a/\sqrt 2)}
\ee
where V is the simulation volume.  
The kernel mass $M_k$ is estimated by the following:
\be 
M_k = \frac{\sum_i N_{hi}M_{hi} f_g}{\sum_i N_{hi}}\,
\label{eqn-kernelmass}
\ee
analogous to using the average halo mass in a bin.  

Poisson errors can be estimated from the Gaussian kernel halo numbers
$N_h$, through use of the formula (\citealt{Heinrich2003}; utilized for
the halo mass function in \citealt{Lukicetal2007}):
\be 
\sigma_{\pm} = \sqrt{N_h + {1 \over 4}} \pm {1 \over 2} \ .
\label{eqn-poissonerr}
\ee 

In what follows, we will show results for all halos with more than 8
particles per halo. This is done for purely diagnostic purposes, since
the poorly resolved halos are strongly affected by particle
shot-noise errors. 

%%%%%%%%%%%%%%%%%%%%%%%%%%%%%%%%%%%%%%%%%%%%%%%%%%%%%%%

\begin{figure}
  \includegraphics[type=pdf,ext=.pdf,read=.pdf,width=.47\textwidth]
{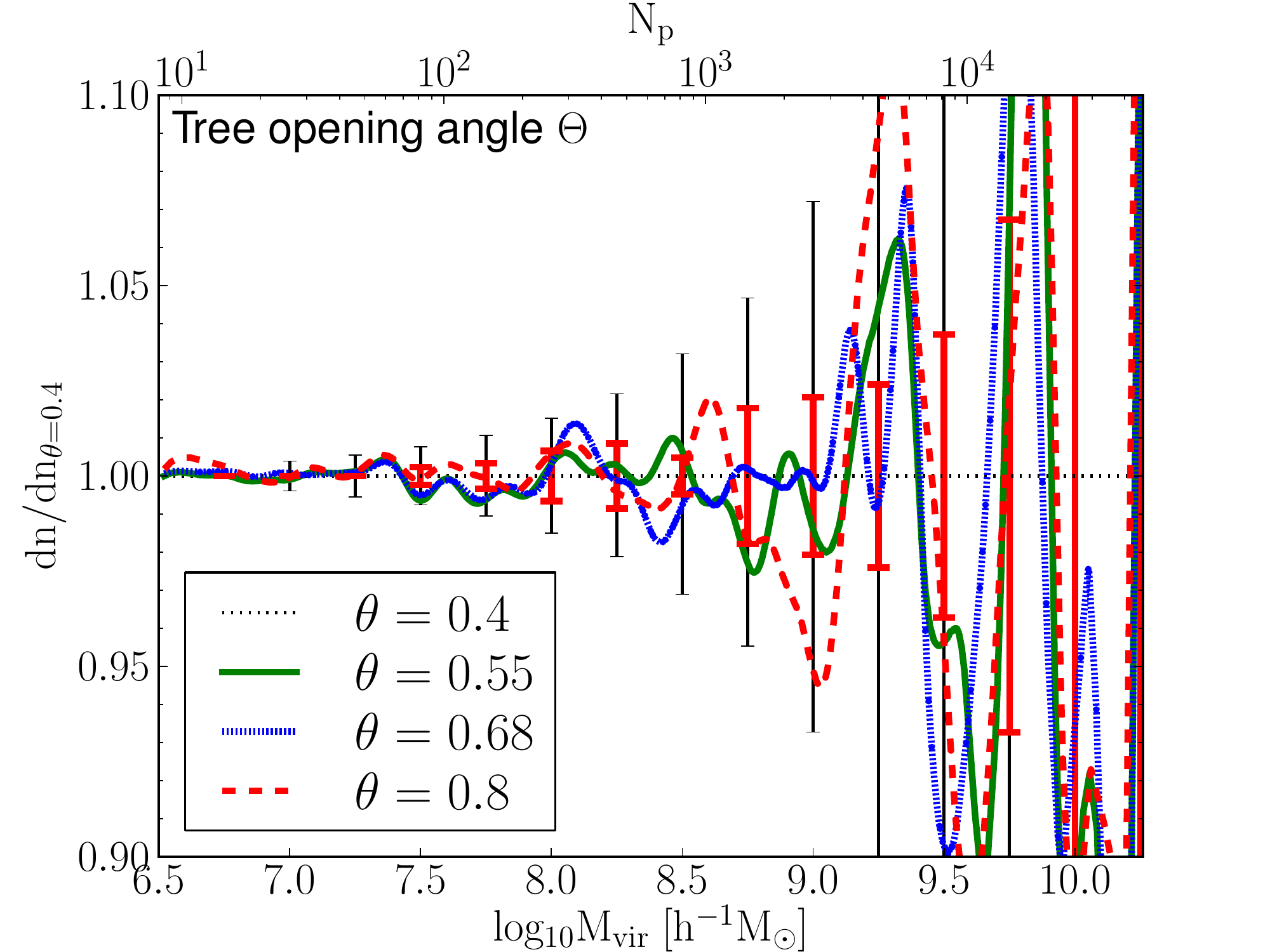}
  \includegraphics[type=pdf,ext=.pdf,read=.pdf,width=.47\textwidth]
{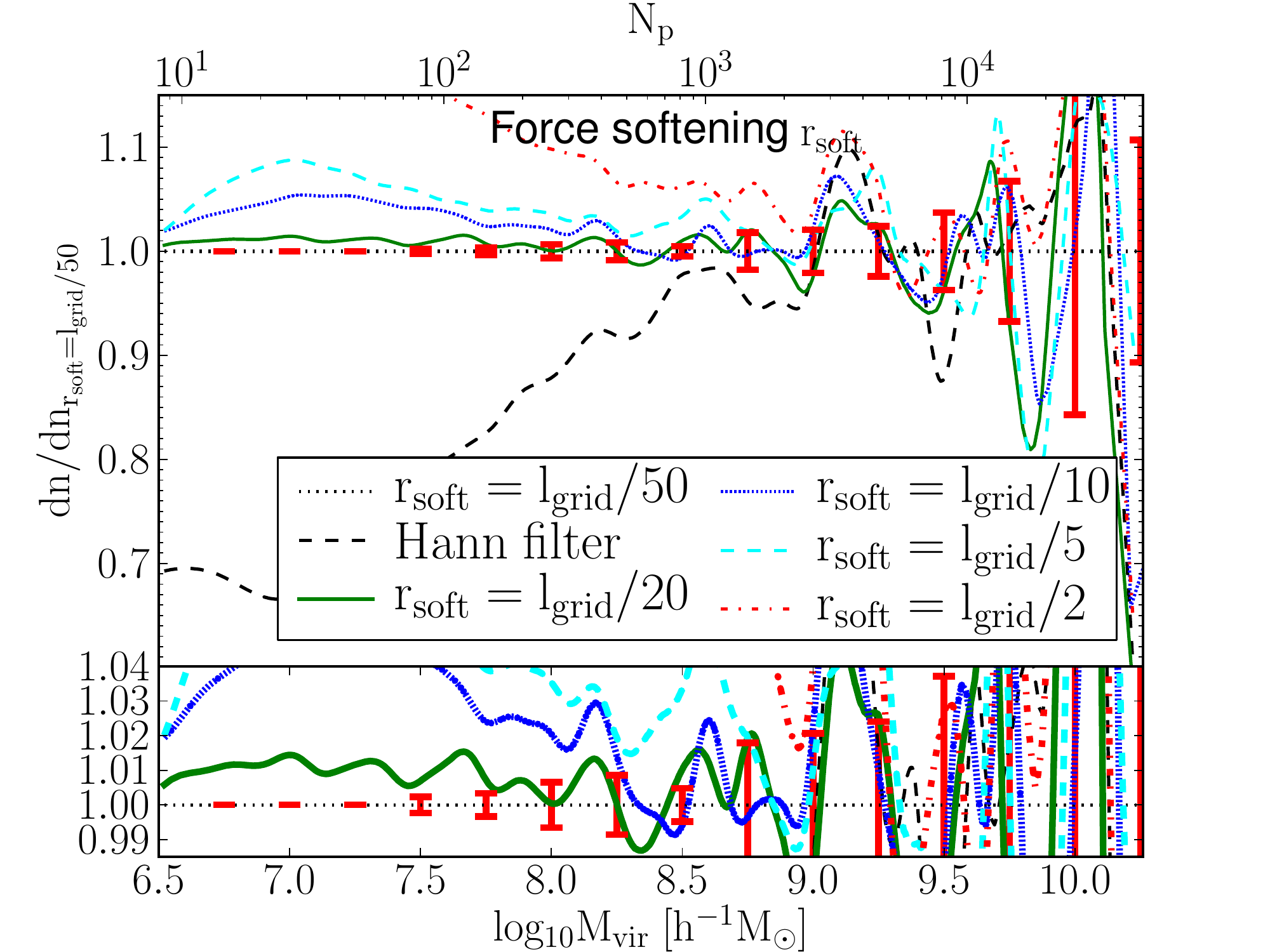}
  \includegraphics[type=pdf,ext=.pdf,read=.pdf,width=.47\textwidth]
{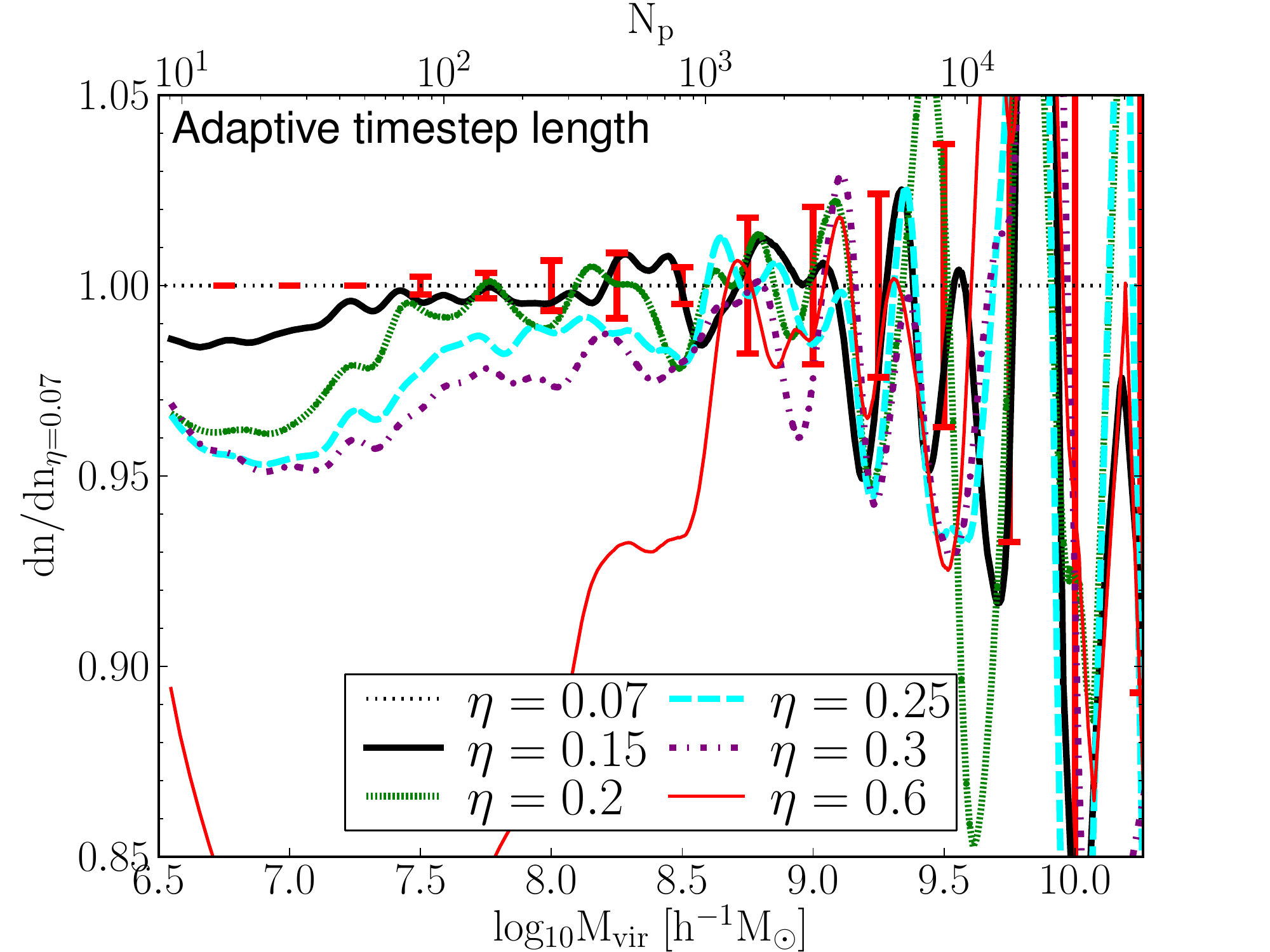}
  \caption{Variation of the halo mass function with simulation
    parameters, relative to the mass function obtained from our
    fiducial simulation, as a function of FoF halo mass. All panels
    show the results from the $L=17.625 \Mpc$, {\tt PKDGRAV}
    simulations at $z=10$. All runs used the same initial realization
    of the density field. Long thin error bars show the Poisson errors
    estimated from the number of halos. Short thick red error bars
    are estimated from the scatter of the mass functions of
    sub-sampled versions of the simulations.  The top, middle and
    bottom panels show the results of variations in tree opening angle
    $\Theta$, force softening $\epsilon$, and adaptive time-step
    parameter $\eta$, respectively.  The black dashed line in the
    middle panel shows the large (and assumedly undesirable) impact of
    application of the Hann anti-aliasing filter to the initial
    density field, on the halo mass function. Percent level
    convergence is apparent for each run parameter for halos larger
    than $\sim100$ particles. }
  \label{fig-runcon}
 \end{figure}

%%%%%%%%%%%%%%%%%%%%%%%%%%%%%%%%%%%%%%%%%%%%%%%%%%%%%%%

\subsection{Tree opening angle $\Theta$}

The top panel of Figure~\ref{fig-runcon} presents the results from our
convergence study of the tree-opening angle parameter $\Theta$, used
in the code {\tt PKDGRAV}. These results are for the case of the
$z=10$, $L=17.625\, \Mpc$ runs. The figure clearly shows that, for
halos with masses $M\le10^9\Msol$, the runs with $\Theta \le 0.68$
are converged at the sub-percent level. For halos with masses
$M\gtrsim10^9\Msol \ (N_p\gtrsim3000)$, the figure shows that scatter
in the mass function begins to dominate our results. This implies that
systematic errors at greater than the percent level cannot be ruled
out for more massive halos, which may be most affected by
tree-related criteria. In this case, the $\Theta=0.8$ run deviates
modestly from the other runs.  Larger values of $\Theta$ have been
shown to cause significant direct force errors \citep{Stadel2001}.

The increased scatter in the mass function for halos with
$M\ge10^9\Msol$ may seem somewhat surprising, given that all of our
simulations had the same initial realization of the density field so
that there is no ``sample variance'' between them.  However, even the
most accurate particle simulation is still essentially a Monte Carlo
representation of a mass distribution. This means that the mass of
each halo has a significant uncertainty, which can at best be equal to
the square root of the number of its particles.  The scatter in the
mass function arises from a convolution of the true halo number counts
with the scatter associated with simulating, sampling, and measuring
the masses of halos (see discussion in \S\ref{sec-halofinding}). Hence
it is non-trivial to determine the true uncertainty.
Fig.~\ref{fig-runcon} shows the expected Poisson errors (long thin
error bars), and one can see that differences between well-converged
runs are at sub-Poisson levels.

For a better estimate of the uncertainty, we create randomly
subsampled 1-in-8 particle volumes from the full simulation snapshot
and measure the $1-\sigma$ scatter between the FoF mass functions of
the subsampled volumes.  In Fig.~\ref{fig-runcon}, the results of this
exercise are denoted by the short, thick, red error bars.  This
scatter tends to be an overestimate of the true uncertainty, since the
scatter in the FoF mass will be larger in the randomly sampled volume
due to the smaller numbers of particles per halo.  For the most
massive halos, this overestimation may become worse due to the
increased effects of halo bridging (or unbridging), which has a large
effect on halo masses and thus on the inferred mass function.  Taking
these issues into account, we estimate that we are sensitive to
percent level systematic shifts in the FoF mass function for halos
with less than $\sim 3000$ particles.

%%%%%%%%%%%%%%%%%%%%%%%%%%%%%%%%%%%%%%%%%%%%%%%%%%%%%%%

\subsection{Force softening $\epsilon$}

The central panel of Figure~\ref{fig-runcon} presents the results from
our convergence study of the force softening parameter $\epsilon$. As
can be seen, the commonly used softening value of $\epsilon=\lm/20$ is
converged at near the percent level over all masses
($\lm=L/N^{1/3}$). We also see that the low-mass end of the mass
function is very sensitive to the choice of $\epsilon$. Halos with
$N\gtrsim 1000$ particles appear only weakly dependent on $\epsilon$,
provided $\epsilon \lesssim \lm/10$ and within our statistical
limitations.

For the cases where $\epsilon\gtrsim\lm/10$, forces do not become
Newtonian until beyond the FoF linking length of $\lm/5$.
Interestingly, for these large softening lengths, we find that the
halo abundances at a fixed mass are increased. One possible
explanation for this effect is that the lower central densities of
the heavily softened halo profiles \citep[\eg]{Mooreetal1998b,FukushigeMakino2001,Poweretal2003,Reedetal2005c,Tinkeretal2008} lead to higher densities near the
outer edges of halos.
This increased outer-shell density, implies
larger FoF masses as more particles are linked to
the outer layers \citep{Bhattacharyaetal2011}; (see \S\ref{sec-rhoeps}). 
For the smallest halos with few particles, the virial
radii and softening are of similar order.  The minimum resolved halo 
mass thus increases as softening increases \citep[\eg]{Lukicetal2007}.
These issues have implications for a
common running mode for cosmological simulations where computational
speed-up at high redshifts is achieved by using a ``physical
softening'', wherein the comoving softening length is scaled by the
expansion factor, with a typical maximum of ${\rm f_{soft~max}\sim
  10 \epsilon}$.  Effectively, in this mode, the
softening is frozen in physical coordinates at scale factor ${\rm a =
  1/f_{soft~max}}$.  Our tests suggest that allowing a comoving
softening larger than $\lm/20$ at high redshifts leads to significant
numerical error for the early forming halos, which is likely to
affect high density regions at late times.

Finally, the black dashed line in the central panel of
\Fig{fig-runcon} shows the results of applying the Hann
anti-aliasing filter. Whilst it may help to correct errors due to the
an-isotropic lattice structure, it introduces a $\sim30\%$ suppression
in the number of lower-mass halos relative to the unfiltered initial
conditions run and relative to the expected nearly power-law behavior
of the mass function predicted from theory \citep{Bondetal1991}. Hann
filtered and unfiltered runs only agree at the percent level for
halos with $N\gtrsim3000$ particles.

%%%%%%%%%%%%%%%%%%%%%%%%%%%%%%%%%%%%%%%%%%%%%%%%%%%%%%%

\begin{figure*}
  \centering{
    \includegraphics[type=pdf,ext=.pdf,read=.pdf,width=1.\textwidth]{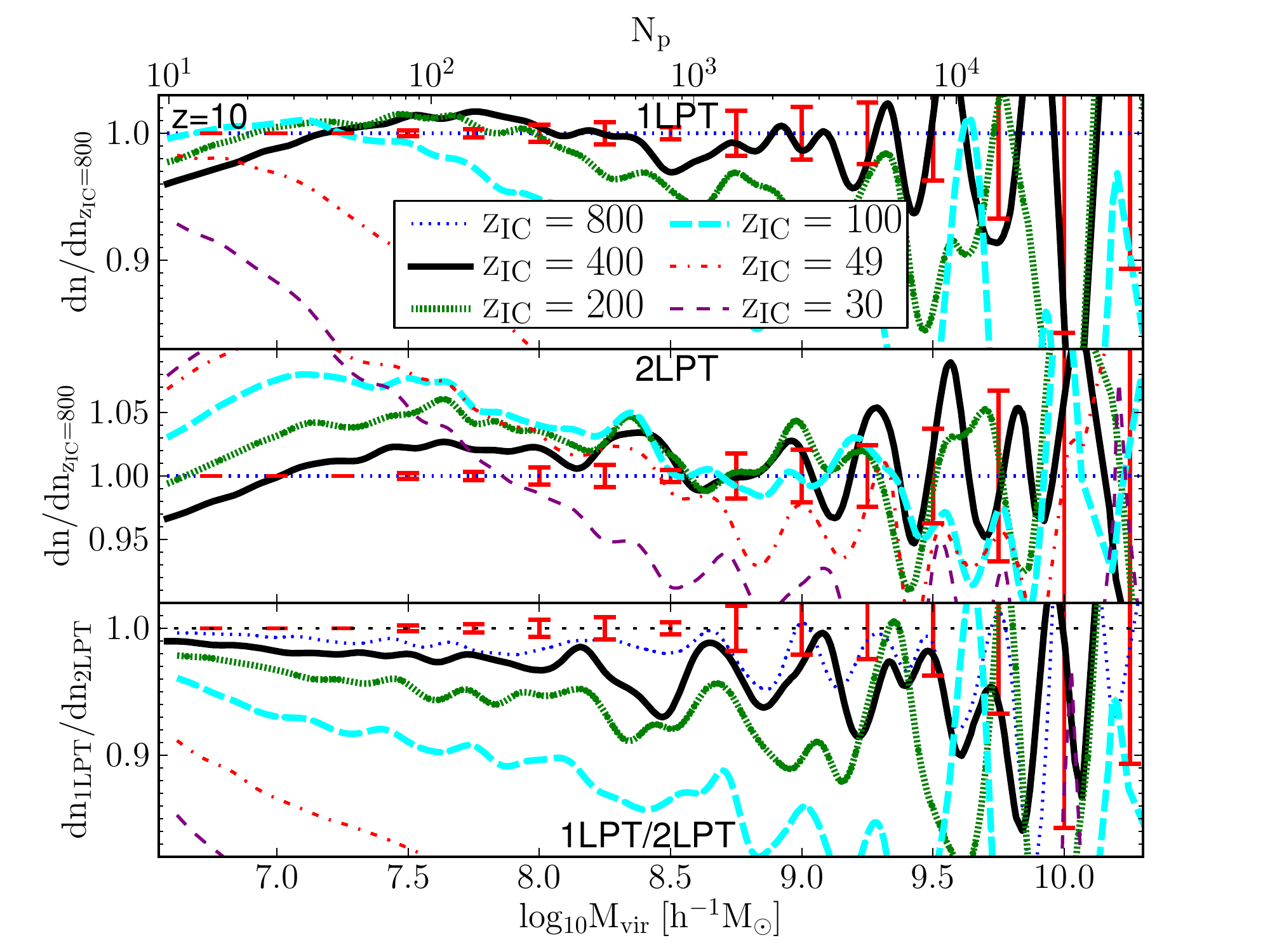}}
  \caption{Variation in the halo mass function as a function of start
    redshift, for 1LPT and 2LPT initial conditions using the $N$-body
    code {\tt PKDGRAV}. {\em Top panel}: results for 1LPT initial
    conditions. {\em Middle panel}: results for 2LPT initial
    conditions. {\em Bottom panel}: the ratio of the 1LPT and 2LPT
    mass functions approaches unity at high initial start
    redshifts. All panels show results of the $L=17.625 \Mpc$ box at
    $z=10.0$.  The 1LPT series displays `false convergence' for $\sim 100$
    particle halos.  Percent level convergence is met for 2LPT for
    $N_P \simgt 1000$ particles to $z_{\rm i}=400$ or $a_{\rm
      f}/a_{\rm i}\approx 40$.}
  \label{fig-1lpt2lptcomb}
\end{figure*}

%%%%%%%%%%%%%%%%%%%%%%%%%%%%%%%%%%%%%%%%%%%%%%%%%%%%%%%

\subsection{Time-step size $\eta$}

The bottom panel of Figure~\ref{fig-runcon} presents our study of how
variations in the adaptive time stepping parameter $\eta$ affects the
estimated mass functions. The results demonstrate that, for halos with
$M\lesssim10^9\Msol$ (3000 particles), an increase in $\eta$ leads to
a decrease in the abundance of halos. We find that percent level
convergence in the mass function can be achieved with
$\eta\approx0.15$ for all halo masses, or $\eta\approx0.2$ for a 100
particle minimum mass.  This value is consistent with the value of
$\eta=0.2$ found by \citet{Poweretal2003}, who examined the
convergence behaviour of the density profiles of dark matter halos.
This similar converged time-step size is not surprising given that
low-redshift halo centers consist of some of the earliest material to
be bound into halos \citep{Diemandetal2005b}.  For halos with
$N\gtrsim1000$ particles, the mass function converges with a much
longer time-step corresponding to $\eta=0.6$.

Finally, we note that the value $\eta=0.2$ for {\tt PKDGRAV} is
similar to the default size of the adaptive time stepping in {\tt
  Gadget-2}: the parameter ${\tt ErrTolIntAccuracy}=\eta^2/2$ has a
default setting of 0.025, which corresponds to $\eta=0.22$.

%%%%%%%%%%%%%%%%%%%%%%%%%%%%%%%%%%%%%%%%%%%%%%%%%%%%%%%

\section{Mass Function convergence II: Initial conditions}
\label{sec-ic}

\subsection{Results: Small boxes at z=10}

Figure \ref{fig-1lpt2lptcomb} compares the behaviour of the 1LPT and
2LPT initial conditions, as a function of the start redshift $z_{\rm
  i}$, for the small-box simulations at $z=10$ evolved with {\tt
  PKDGRAV}.  The top and middle panels show the ratios of the halo
mass functions for different start redshifts with the halo mass
function obtained from the start redshift $z_{\rm i}=800$, for 1LPT
and 2LPT, respectively. The bottom panel presents the ratio of the
1LPT and 2LPT mass functions for simulations with the same start
redshift.  We see that both the 1LPT and 2LPT initial conditions
converge to yield the same mass function as start redshift
increases. However, the convergence properties of the 1LPT runs are
very slow, whereas the 2LPT runs appear to converge much faster.  

For the case of 2LPT, we notice that percent level convergence can be
achieved for halos with at least $\sim 1000$ particles and in
simulations that have undergone $a_{\rm f}/a_{\rm i}\gtrsim 10$ expansions. 
%up to our statistical limit of halos with $\sim 3000$ particles 
For 1LPT, the $a_{\rm f}/a_{\rm i}=80$ run ($z_{\rm i}=800$) is barely
converged down to $N\sim 1000$.  For halos, with $N\lesssim 1000$
particles, even the 2LPT mass function is poorly converged with
$z_{\rm i}$ for all expansion factors tested. The abundances of small
halos appear to diminish as start redshift is increased; this is
apparent in both the 2LPT and 1LPT tests.  This suggests that
$N\sim1000$ particles represents a minimum halo mass for stability to
$z_{\rm i}$.

A curious coincidental feature of the 1LPT initial condition series is
that small halos appear to be converged at the $\sim2\%$ level by
$z_{\rm i}=200$ (except for the largest masses) and nearly at the $\sim1\%$ level by
$z_{\rm i}=400$ (except for the smallest masses).  The more accurate 2LPT start redshift series
confirms that this 1LPT convergence is an illusion.  With later start
redshift, the increased number of halos due to more accuracy in the
1LPT initial conditions is offset by independent errors that act to
decrease the number of halos, resulting in {\it false convergence}.
This highlights the fact that convergence is a necessary but not
sufficient condition to guarantee accuracy.  Some previous studies
that appeared to show good mass function convergence with early enough
1LPT initialization, such as \citet{Reedetal2003} and \citet{Reedetal2007b} (Fig. A1), among others, likely also
suffered from this false convergence.  Our larger particle numbers and
corresponding better statistical uncertainty, combined with 2LPT
comparisons, allow us to make more robust conclusions.

%%%%%%%%%%%%%%%%%%%%%%%%%%%%%%%%%%%%%%%%%%%%%%%%%%%%%%%

\begin{figure}
  \centering{
    \includegraphics[type=pdf,ext=.pdf,read=.pdf,width=.52\textwidth]{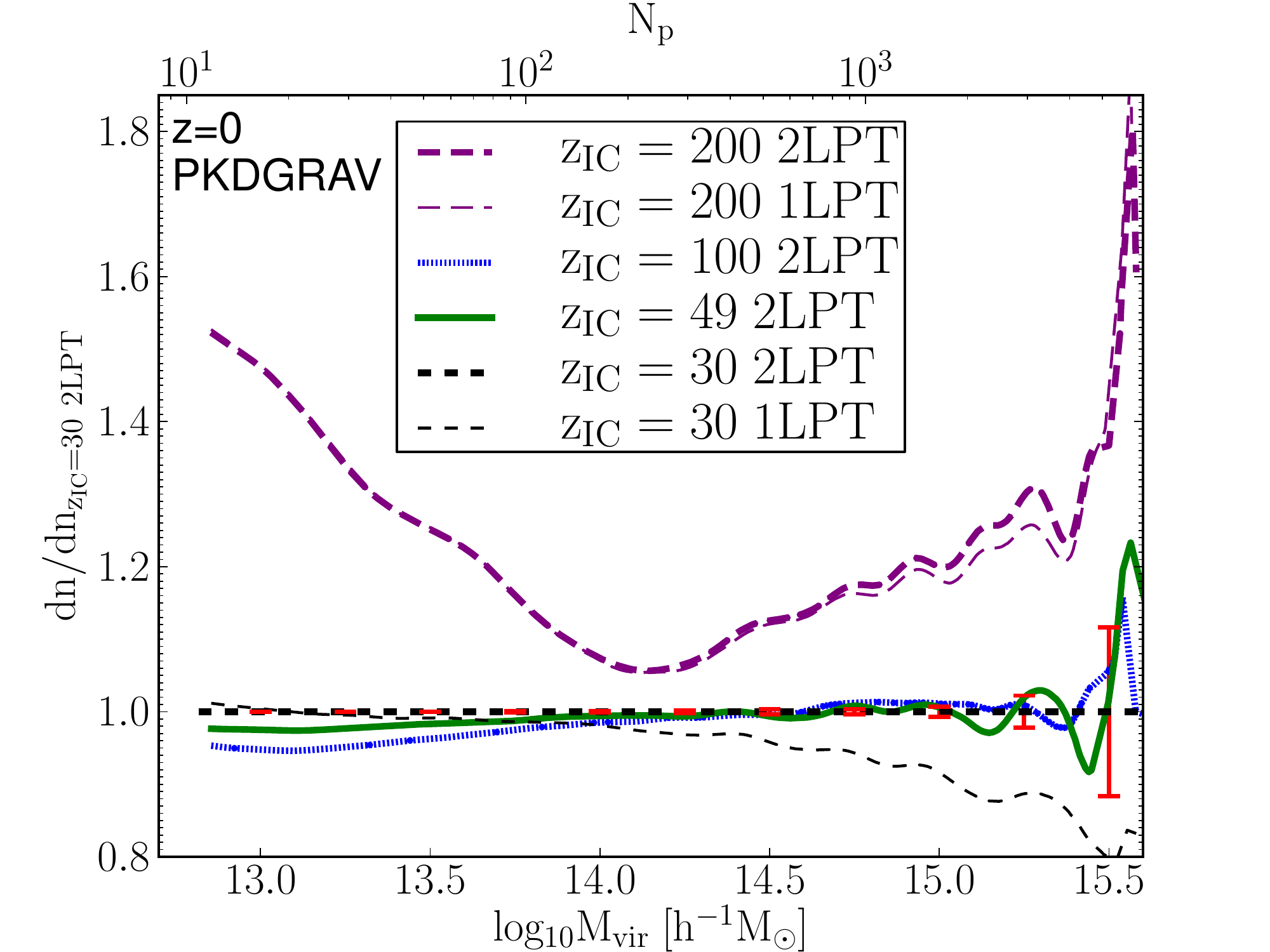}}
  \caption{Mass functions from the simulations started with 1LPT and
    2LPT initial conditions in the $L=2\Gpc$ boxes at ${z=0}$, relative
    to the simulation started at $z_{\rm i}=30$ with 2LPT initial
    conditions.  All simulations here were run with {\tt PKDGRAV}.
    Percent level convergence with 2LPT is found between $z_{\rm
      i}=30$ and $z_{\rm i}=49$ runs at $\sim 1000$ particles.  
    Extremely early starts ( $a_{\rm f}/a_{\rm i} \gtrsim 100$) lead
    to serious errors.}
  \label{fig-1lpt2lptbig1}
\end{figure}

%%%%%%%%%%%%%%%%%%%%%%%%%%%%%%%%%%%%%%%%%%%%%%%%%%%%%%%

\subsection{Results: Large boxes at z=0}
\label{sec-largeboxes}

In order to confirm that the convergence behavior of the small box
simulations at $z=10$ can be applied to the cluster regime at lower
redshifts, we present the results from tests run to $z=0$ in the
$L=2\Gpc$ boxes.

Figure~\ref{fig-1lpt2lptbig1} shows the results from the 1LPT and 2LPT
$z_{\rm i}$ convergence simulations run with the code {\tt
  PKDGRAV}. As for the case of the small-box simulations at $z=10$, we
see that low-mass halos are missing with high start redshifts
$z\gtrsim49$; the $z=0$ ``pivot'' mass-scale, below which convergence
is poor, is somewhat smaller at $N\sim 300$ particles.  For larger
halos, the 2LPT mass function is well-converged so long as $z_{\rm i}
\simlt 49$; the $z_{\rm i}=100$ curve deviates from the $z_{\rm i}=30$
curve at just over the percent level at this mass scale, so is
marginally statistically consistent at the percent level.  A striking
feature here is that when a high enough start redshift is used that
Zel'Dovich and 2LPT initial conditions are converged, $z_{\rm i}=200$,
serious errors are present in the $z=0$ mass function.  A likely
explanation is that with such a high $z_{\rm i}$, cosmological
perturbation amplitudes becomes smaller than the effective amplitude
of spurious numerical perturbations.  In this $z_{\rm i}=200$ run,
spurious halos begin to form at very early times, initially dominated
by 8 particle structures; visual inspection reveals that the spurious
halos are aligned with the initial grid of particles.  The effects of
these early spurious halos lead to the over-abundance of halos at
$z=0$.  This underscores the point that 2LPT initial conditions are
preferable because they allow one to start at lower redshift where
numerical errors are more controllable.

We note that pure particle mesh codes may perform better with high
redshift starts because the PM technique is well-suited to following
low amplitude linear perturbations.  A tree code (and also a tree-PM
code), however, is subject to force errors that may accumulate over
time, even if they are small, because these errors are correlated with
the tree structure.  Tree code force errors could thus seed spurious
structures that dominate over real cosmological perturbations when
start redshift is very high (and initial cosmological perturbations
are very low).  Further, the accumulated errors would be expected to
worsen if time-step length is decreased.  The PM code, although it may
perform better at early times, is limited in spatial resolution by the
mesh size, which is typically much larger than the softening scale of
a tree (or tree-PM) code, making it non-ideal for modeling the
internal properties of halos.

%%%%%%%%%%%%%%%%%%%%%%%%%%%%%%%%%%%%%%%%%%%%%%%%%%%%%%%

\begin{figure}
  \includegraphics[type=pdf,ext=.pdf,read=.pdf,width=.52\textwidth]{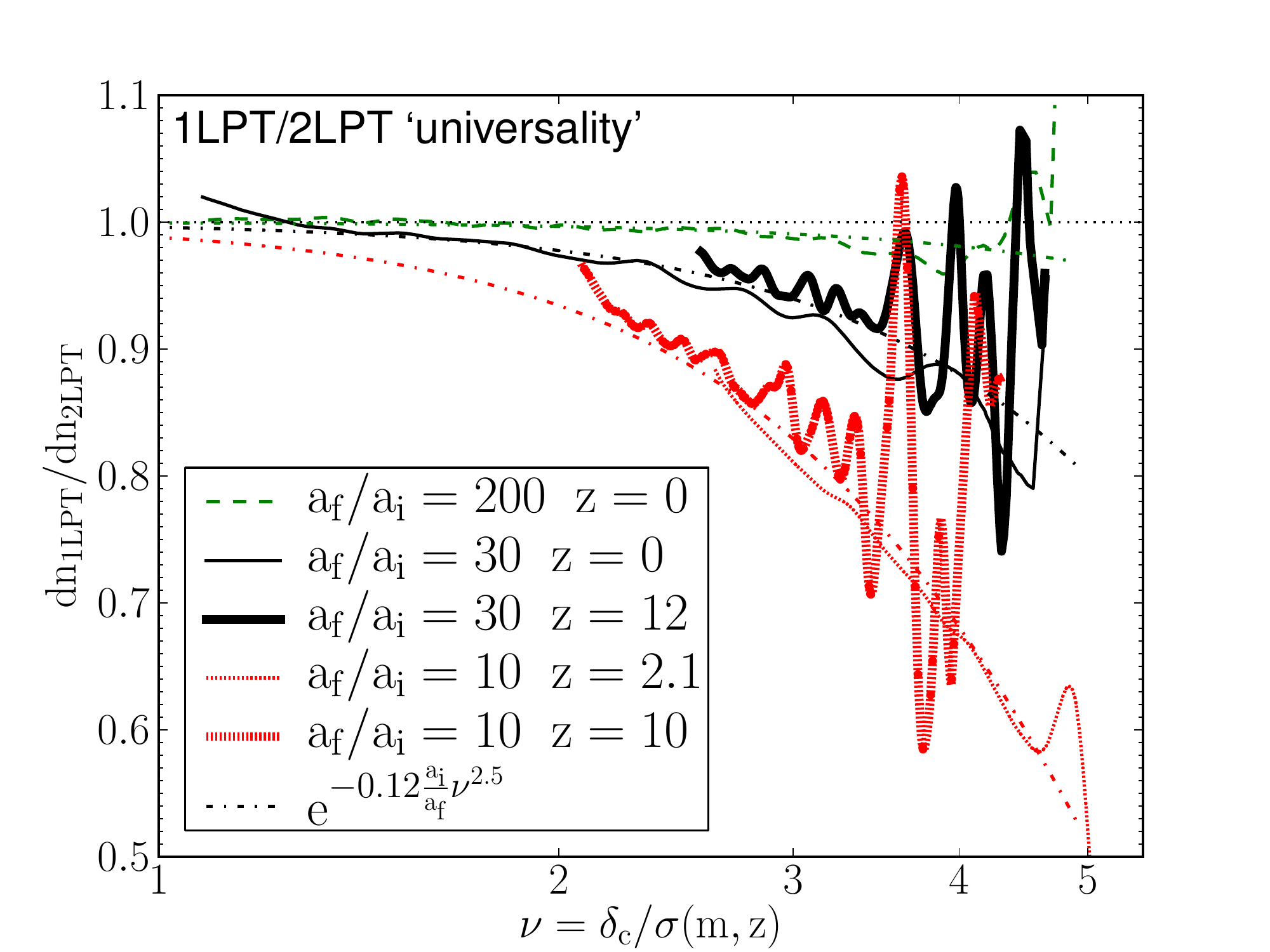}
  \caption{Ratio of 1LPT mass functions with those obtained from 2LPT
    initial conditions as a function of the equivalent halo peak
    height $\nu$. The thick solid black and thick hatched red line
    show results from the $L=17.625\Mpc$ boxes at ${\rm z \ge 10}$,
    and the thin green dashed, solid black and hatched red show
    results from the $L=2\Gpc$ boxes. All results here are for {\tt
      PKDGRAV}.  The ratio of 1LPT/2LPT mass functions is `universal'
    in that it depends mainly upon $a_{\rm f}/a_{\rm i}$ and $\nu$,
    independent of the specific halo mass range and redshift, and fit by the dot-dashed line.}
  \label{fig-1lpt2lptuni}
 \end{figure}

%%%%%%%%%%%%%%%%%%%%%%%%%%%%%%%%%%%%%%%%%%%%%%%%%%%%%%%

\subsection{Transformation to universality}

It has been noted that when halo masses are translated into equivalent
`peak-height', i.e., where $M\rightarrow \nu(M,a)$, then the mass
function takes on a `universal form' \citep{ShethTormen1999}. The peak
height is defined through the relation, $\nu=\delta_c/\sigma(M,z)$,
where $\delta_c=1.686$ is the present day linearly extrapolated
over-density threshold for collapse in the spherical collapse model,
and where $\sigma(M,z)$ is the variance of matter fluctuations on mass
scale $M=4\pi R^3\rhob/3$. Note also that owing to the fact that
$\sigma(M,a)\propto D(a)$, $\nu\propto D^{-1}(a)$, where $D(a)$ is the
linear theory growth factor.

In Figure~\ref{fig-1lpt2lptuni} we present the ratio of the 1LPT with
the the 2LPT mass functions for the small box at $z=10$ and the big
box at $z=0$ as a function of $\nu$. We see that the 1LPT mass
function error appears to be relatively independent of mass and
redshift for equal values of $\nu$.  This universal behavior is
expected in Press-Schechter formalism wherein mean halo formation time
depends only upon $\nu$, and $\delta_c$ is independent of redshift.
We present a fit to the ratio of the 1LPT to 2LPT mass function for 
our range of data $1 \simlt \nu \simlt 5$,
\be 
{\rm dn_{1LPT}/dn_{2LPT}= e^{-0.12 {a_i \over a_f} \nu^{2.5}}}.
\label{eqn-unifit}
\ee 
Percent level convergence between 1LPT and 2LPT initial conditions at
scales relevant for the cluster mass function is not achieved until
extremely early starts at $a_{\rm f}/a_{\rm i} \sim 200$. However,
such early start redshifts lead to very small initial density
perturbations, which through the relative increase of numerical
errors, preclude the codes that we tested from being able to
accurately follow the growth of structure.

%%%%%%%%%%%%%%%%%%%%%%%%%%%%%%%%%%%%%%%%%%%%%%%%%%%%%%%

\subsection{Comparison of {\tt PKDGRAV} with {\tt Gadget-2}}

The quest for obtaining mass function predictions accurate to within
one percent requires different $N$-body simulation codes to provide
consistent results at this level. Of course, if results disagree at
$\gtrsim1\%$ for any two codes, then one would need at least three
independent $N$-body codes to break the degeneracy and so decide which
results were correct. Having said that, we now compare the initial
condition convergence series of simulations obtained using {\tt
  PKDGRAV} with results obtained from the widely used $N$-body code
{\tt Gadget-2}.  Note that we have made no attempt to find `optimal'
parameters for {\tt Gadget-2}, but instead we have adopted some rather
generic choices for these.  The full list of simulations that we have
performed with {\tt Gadget-2}, including the exact choices for run
parameters, are presented in Table~\ref{table-sims}.

\begin{figure*}
  \centering{
    \includegraphics[type=pdf,ext=.pdf,read=.pdf,width=.78\textwidth]{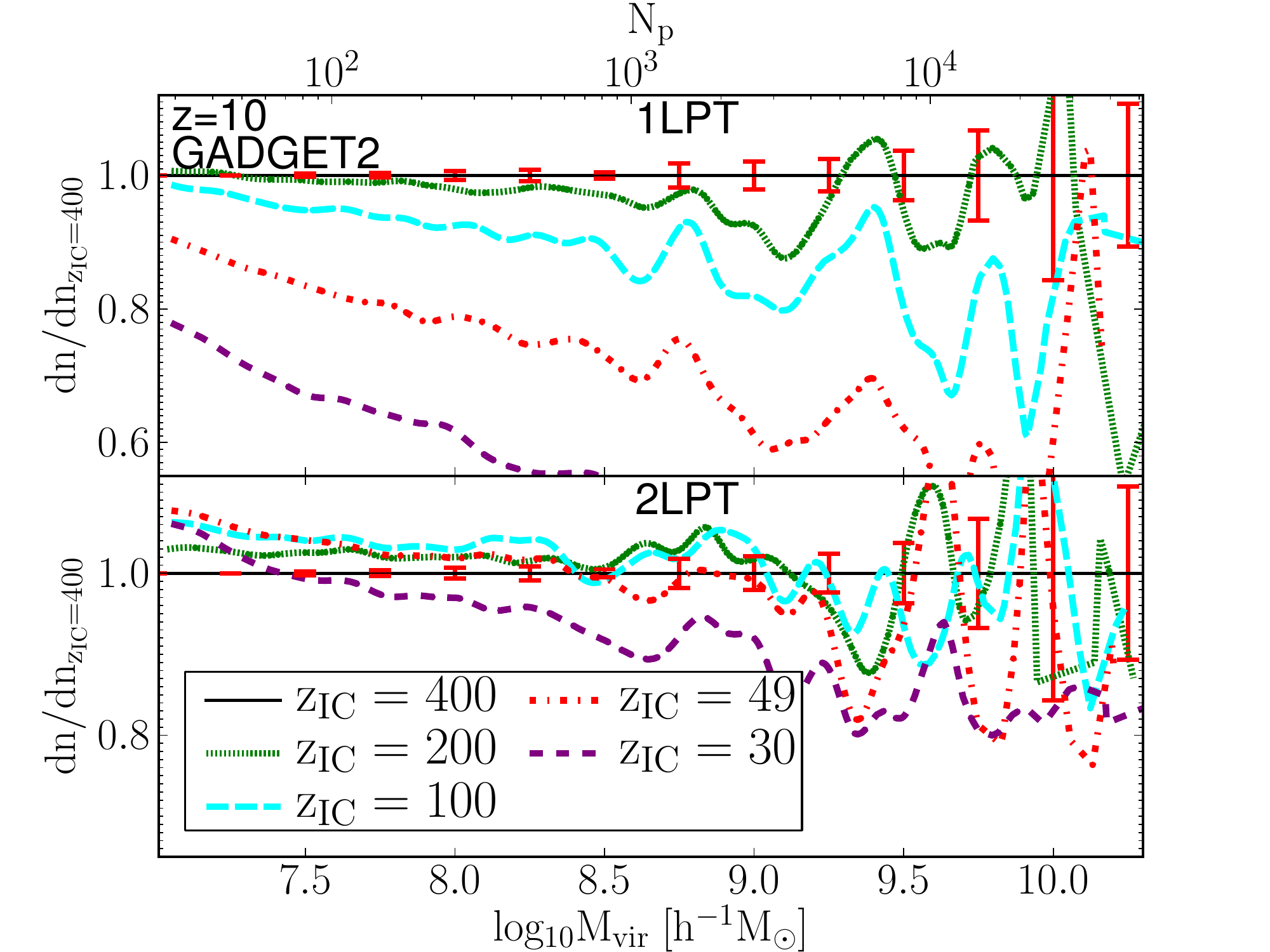}} 
  \caption{Variation in the halo mass function as a function of start
    redshift, for 1LPT and 2LPT initial conditions using the $N$-body
    code {\tt GADGET-2}. {\em Top panel}: results for simulations
    started with 1LPT initial conditions. {\em Bottom panel}: results
    for simulations started with 2LPT initial conditions. All panels
    are for the $L=17.625 \Mpc$ box at $z=10.0$.  Convergence of {\tt
      GADGET-2} runs with initial conditions are very similar that
    found for {\tt PKDGRAV} (Fig.~\ref{fig-1lpt2lptcomb}).}
    \label{fig-1lpt2lptsmallgadg}
 \end{figure*}

In Figure~\ref{fig-1lpt2lptsmallgadg} we compare the 1LPT and 2LPT
initial conditions as a function of $z_{\rm i}$, but this time using
{\tt Gadget-2}, plotted here down to the limit $N=20$ particles. We
find that the results exhibit almost identical behaviour to those
obtained from the {\tt PKDGRAV} runs. The small difference is that the
suppression of the mass function at low masses with increasing start
redshift, apparent in 2LPT runs for halos with fewer than $N\sim 1000$
particles, is slightly milder with {\tt Gadget-2}. This appears to
enhance the effect of ``false convergence'' of the 1LPT {\tt Gadget-2}
simulations.

\begin{figure*}
  \centering{
  \includegraphics[type=pdf,ext=.pdf,read=.pdf,width=.78\textwidth]{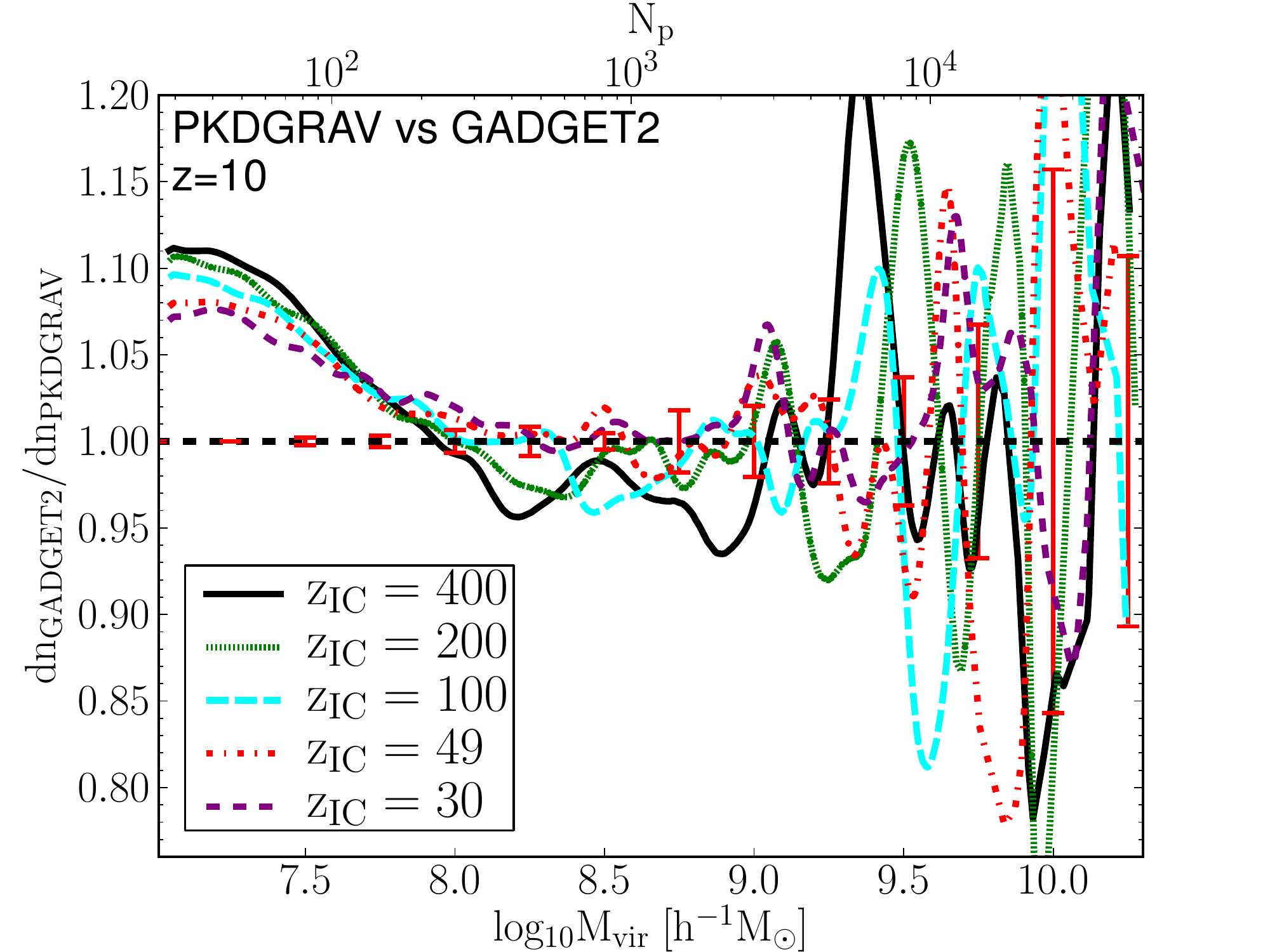}}
  \caption{$z=10.$ 2LPT.  Mass functions from {\tt PKDGRAV} and
    compared with {\tt Gadget-2}.  Some differences are present at
    small scales but the relative offset is independent of start
    redshift, providing some confirmation of our initial condition
    convergence criteria.  We have verified the {\tt PKDGRAV}, but not
    the {\tt Gadget-2} run parameters for percent-level
    self-convergence.  }
  \label{fig-gadgetcomp}
 \end{figure*}

Figure~\ref{fig-gadgetcomp} shows the ratio of the mass functions
obtained from {\tt Gadget-2} with those obtained from {\tt PKDGRAV}.
Note that we used identical initial conditions in all cases.  We find
that {\tt Gadget-2} systematically produces up to a $10\%$ higher mass
function for low-mass halos ($N\lesssim200$) than {\tt PKDGRAV}. This
excess abundance slightly increases with increasing start redshift.
We note that the differing mass functions could be a result of
differing halo structure, which could lead to systematic differences
in FoF masses, and does not necessarily mean the codes are truly
producing different numbers of halos.  We also note that {\tt
  Gadget-2} appears to have several percent fewer halos at $N\sim
1000$ particles, for the highest start redshifts.

\begin{figure}
  \centering{
    \includegraphics[type=pdf,ext=.pdf,read=.pdf,width=.52\textwidth]{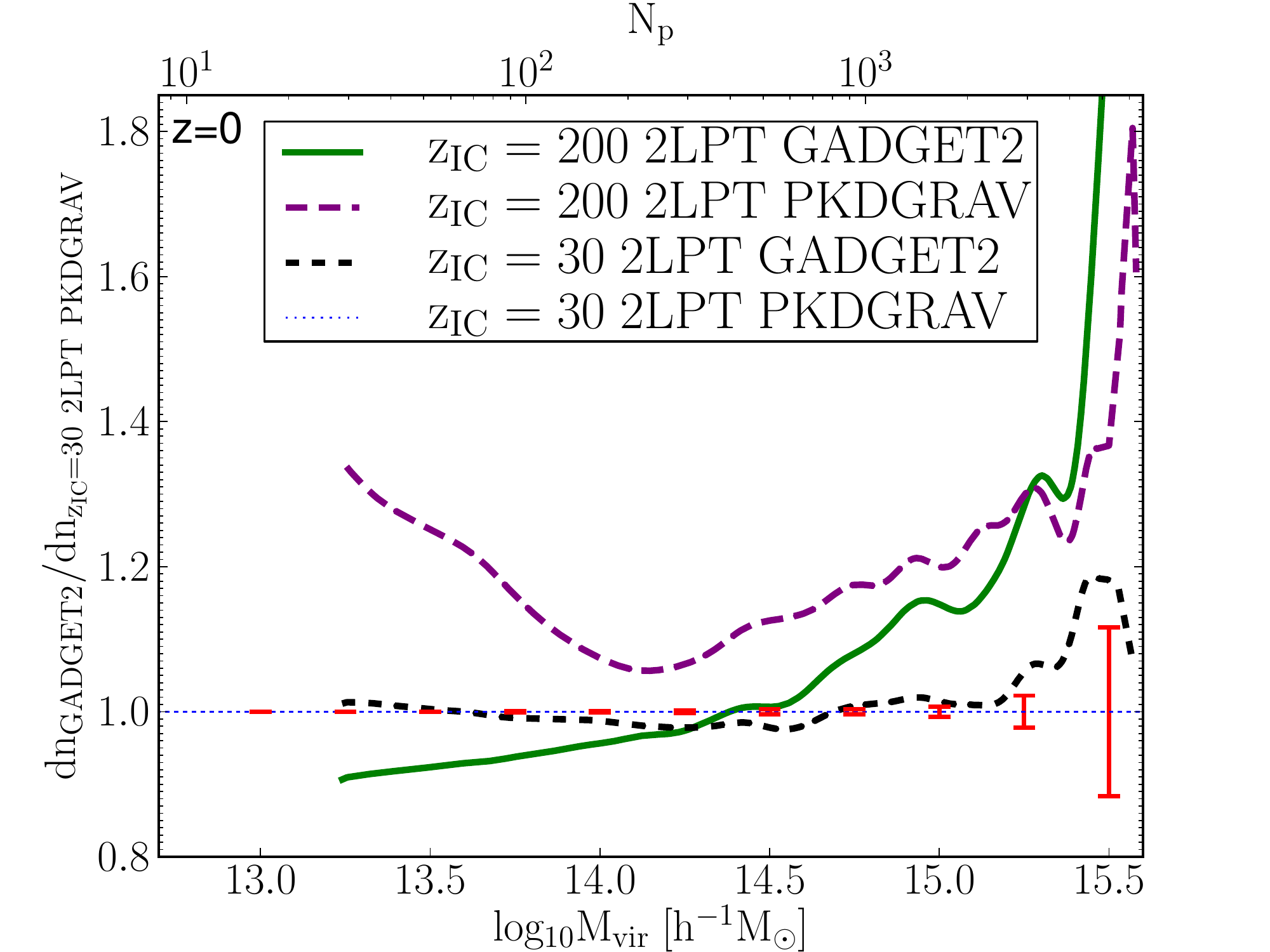}}
  \caption{Mass functions from {\tt Gadget-2} compared with {\tt
      PKDGRAV} in the $L=2\Gpc$ box.  The agreement with {\tt
      Gadget-2} using relatively standard run parameters is at the
    couple percent level with $z_{\rm i}=30$.  Extremely early starts
    ( $a_{\rm f}/a_{\rm i} \gtrsim 100$) lead to serious errors
    independent of the simulation code used.}
  \label{fig-1lpt2lptbiggadg}
\end{figure}

Figure~\ref{fig-1lpt2lptbiggadg} compares 2LPT mass functions from
{\tt Gadget-2} and {\tt PKDGRAV} in the large box at $z=0$.  This
figure shows that when the lower redshift start is used, $a_{\rm
  f}/a_{\rm i} = 30$, the two codes generally agree within $2\%$.
However, when high redshift starts are used, $a_{\rm f}/a_{\rm i} \sim
200$, the codes diverge from each other and from the true answer --
recall Figure~\ref{fig-1lpt2lptbig1} where we showed that the lower
redshift start is converged in {\tt PKDGRAV} runs.

This code comparison shows that there is a weak systematic shift with
start redshift between the codes.  However, the convergence behavior
of {\tt Gadget-2} with start redshift and of 1LPT versus 2LPT still
provides useful verification of the {\tt PKDGRAV} initial condition
tests.  Ultimately, a robust comparison of absolute accuracy between
codes would require that run parameters for each code are run at
self-converged values.  The code differences are consistent with the
level of agreement found between these same codes in
\citet{Heitmannetal2008} when considering our improved statistics and
resolution. Further investigation is warranted to reveal whether the
differences between the two codes is caused by inherent differences
between the TreePM and the pure tree method or whether they are
instead due to the use of non-ideal run parameters in {\tt Gadget-2}.
This is beyond the scope of this paper and we shall reserve a wider
study for future work.

%%%%%%%%%%%%%%%%%%%%%%%%%%%%%%%%%%%%%%%%%%%%%%%%%%%%%%%

%\FloatBarrier
%\clearpage

%%%%%%%%%%%%%%%%%%%%%%%%%%%%%%%%%%%%%%%%%%%%%%%%%%%%%%%

\section{convergence of other properties}
\label{sec-conother}

In this section we consider the sensitivity of other dark matter
statistics: to the adopted simulation parameters; to whether we employ
1LPT or 2LPT initial conditions; and to the adopted start redshift.
We shall restrict our exploration to the mass power spectrum and the
1-point Probability Distribution Function (PDF) of dark matter
density, as additional diagnostics for determining simulation
accuracy.

%%%%%%%%%%%%%%%%%%%%%%%%%%%%%%%%%%%%%%%%%%%%%%%%%%%%%%%

\subsection{Mass power spectra}

For a finite cubical patch of the Universe, the matter power spectrum
is defined to be:
\be \left<\delta_{\bk_1}\delta^*_{\bk_2}\right>
\equiv P(|k_1|)\delta^{K}_{\bf k_1,k_2} /V_{\mu}\ee
where $V_{\mu}$ is the volume of the patch and $\delta_{\bk}$ is the
discrete Fourier series expansion of the density field. For equal mass
dark matter particles, the discrete representation of the Fourier
space density field can be written \citep{Peebles1980}:
\ba
\delta(\bx) & = & \frac{V_{\mu}}{N}\sum_{i}^{N} \delta^{D}(\bx-\bx_i) -1 \Leftrightarrow  \nn \\
\delta(\bk) & = & \frac{1}{V_{\mu}}\int \dx \delta(\bx) e^{i\bk\cdot\bx} \ ,
\ea
where $N$ is the number of particles. The matter power spectra were
estimated for each simulation using the standard Fourier based methods
\citep{Smithetal2003,Jing2005,Smithetal2008b}: particles and halo
centers were interpolated onto a $1024^3$ cubical mesh, using the CIC
algorithm \citep{HockneyEastwood1988}; the Fast Fourier Transform of
the discrete mesh was computed using the FFTW libraries; the power in
each Fourier mode was estimated and then corrected for the CIC charge
assignment; these estimates were then bin averaged in spherical shells
of logarithmic thickness.

Before we proceed to the results, we note that it is not necessarily
the case that a simulation that yields a $\lesssim1\%$ accurate halo
mass function should also yield a $\lesssim1\%$ accurate matter power
spectrum, and {\it vice-versa}.  Different requirements for simulation
parameters are possible because the mass range in our mass functions
only involves a small fraction of the total mass in the simulation.
And because, our estimates of the measured power spectra do not extend
to scales as small as the virial radius of the smallest halos
considered.

%%%%%%%%%%%%%%%%%%%%%%%%%%%%%%%%%%%%%%%%%%%%%%%%%%%%%%%

\begin{figure}
  \centering{
    \includegraphics[type=pdf,ext=.pdf,read=.pdf,height=0.38\textwidth,width=.5\textwidth]{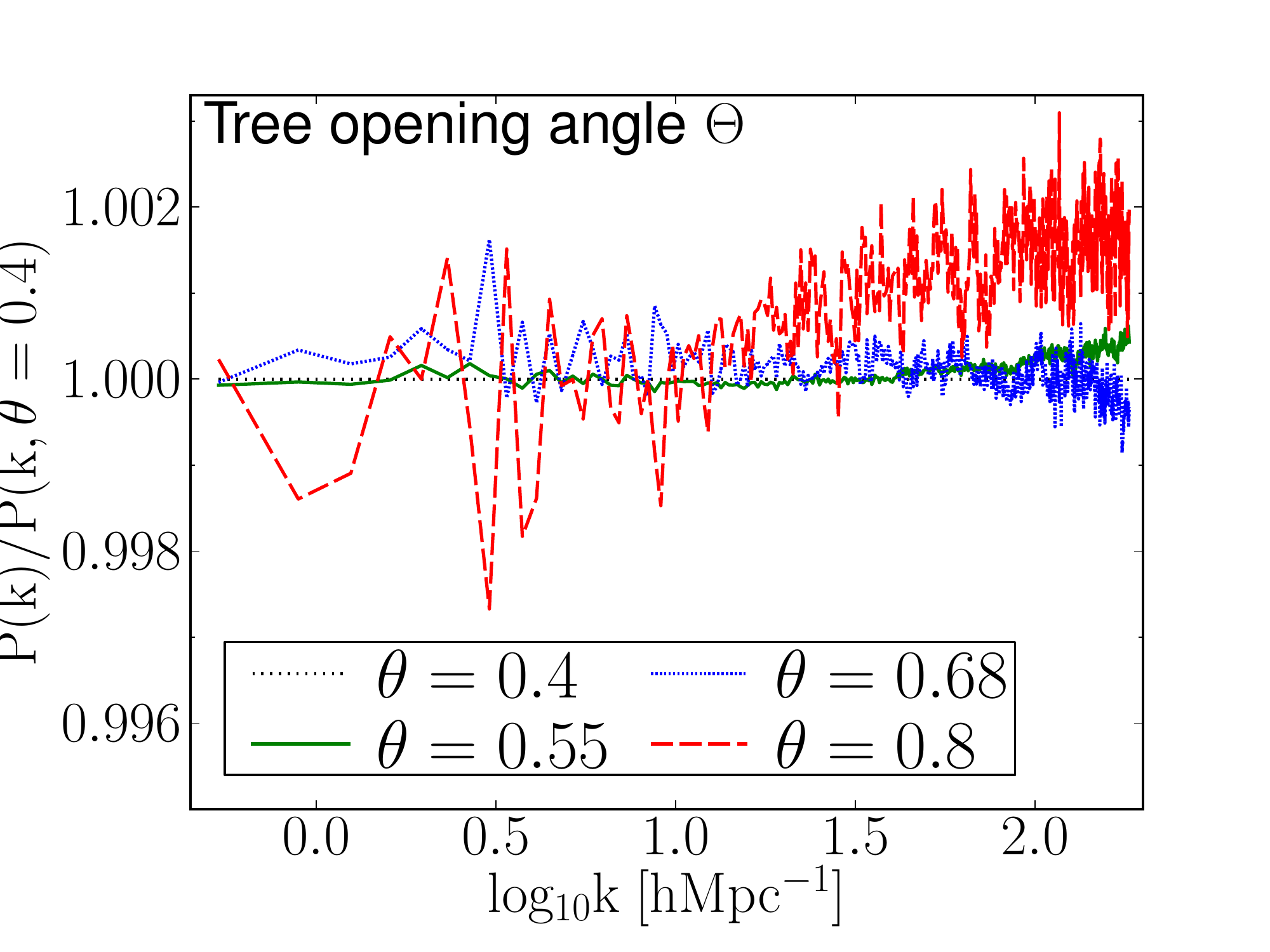}}
  \centering{
    \includegraphics[type=pdf,ext=.pdf,read=.pdf,height=0.38\textwidth,width=.5\textwidth]{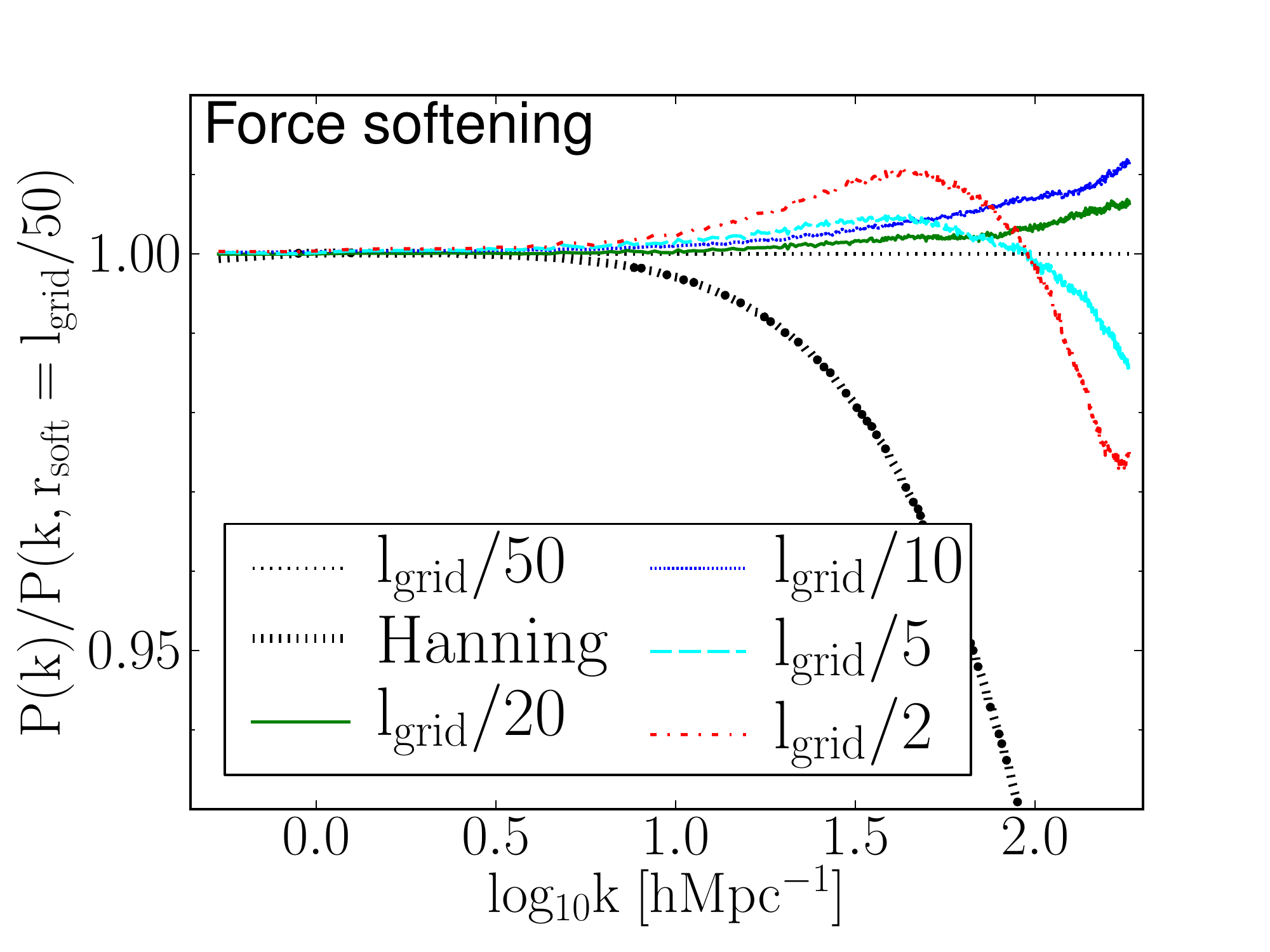}}
  \centering{
    \includegraphics[type=pdf,ext=.pdf,read=.pdf,height=0.38\textwidth,width=.5\textwidth]{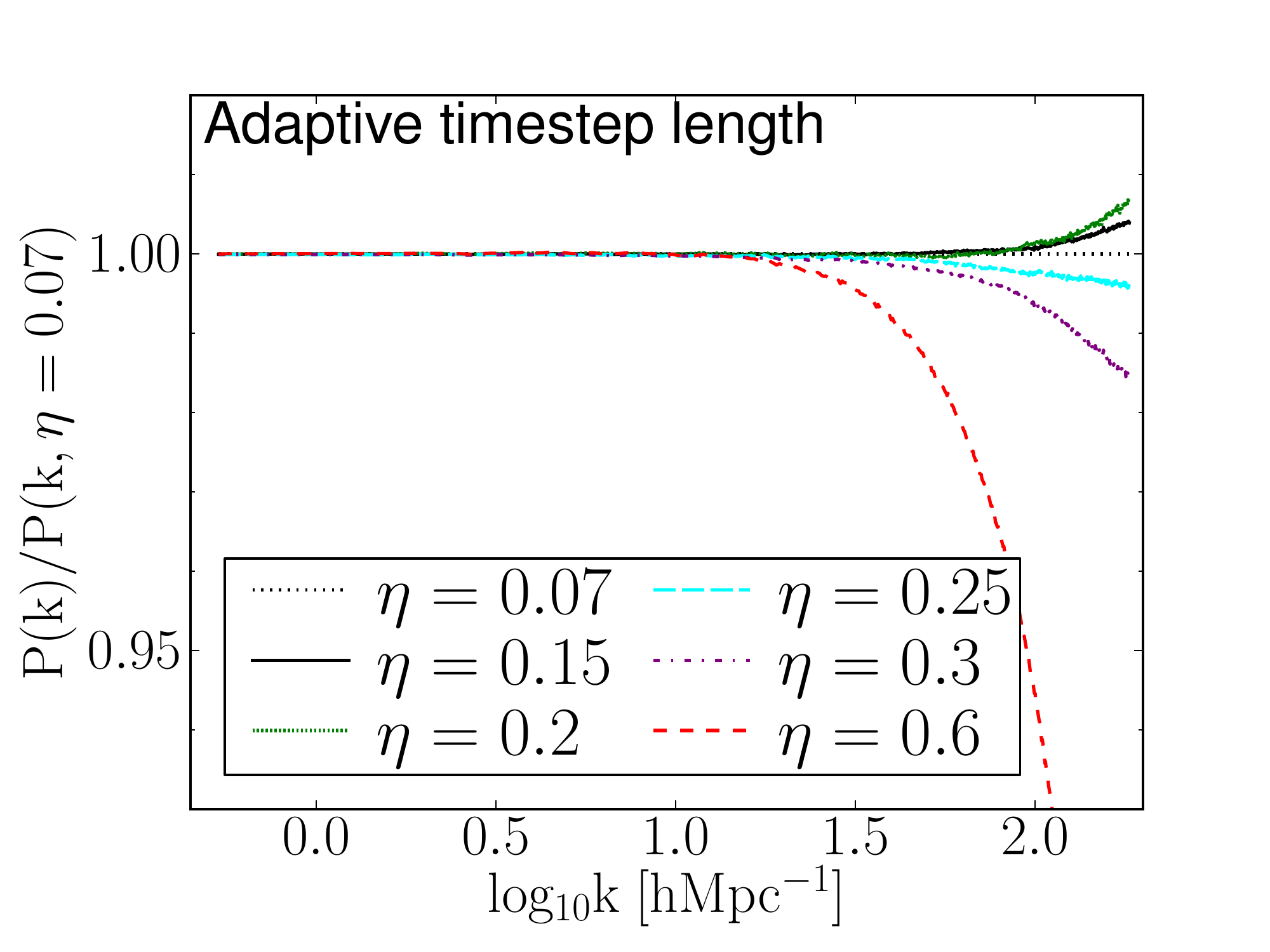}}
  \caption{Relative variation of the dark matter power spectra with
    the simulation run parameters for the $N$-body code {\tt
      PKDGRAV}. Top panel: variation with respect to the tree-opening
    angle $\Theta$. Middle panel: variation with respect to the force
    softening parameter $\epsilon$. Bottom panel: variation with
    respect to the time-step parameter $\eta$. All panels show the
    $L=17.625$ box at $z=10$. Percent level convergence is seen for
    each run parameter. }
  \label{fig-powz10run}
\end{figure}

%%%%%%%%%%%%%%%%%%%%%%%%%%%%%%%%%%%%%%%%%%%%%%%%%%%%%%%

\begin{figure}
  \centering{
    \includegraphics[type=pdf,ext=.pdf,read=.pdf,height=0.38\textwidth,width=.5\textwidth]{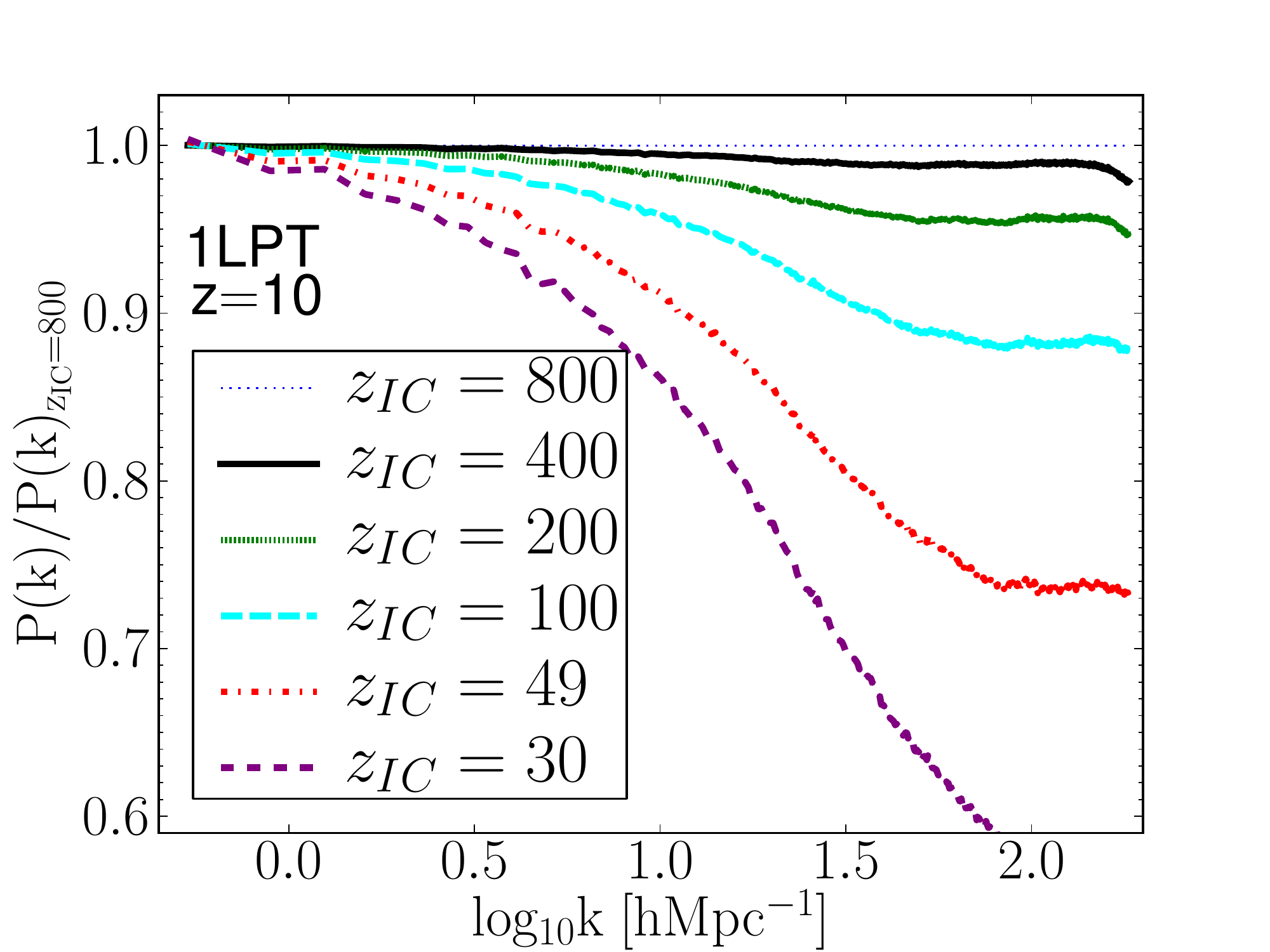}}
  \centering{
    \includegraphics[type=pdf,ext=.pdf,read=.pdf,height=0.38\textwidth,width=.5\textwidth]{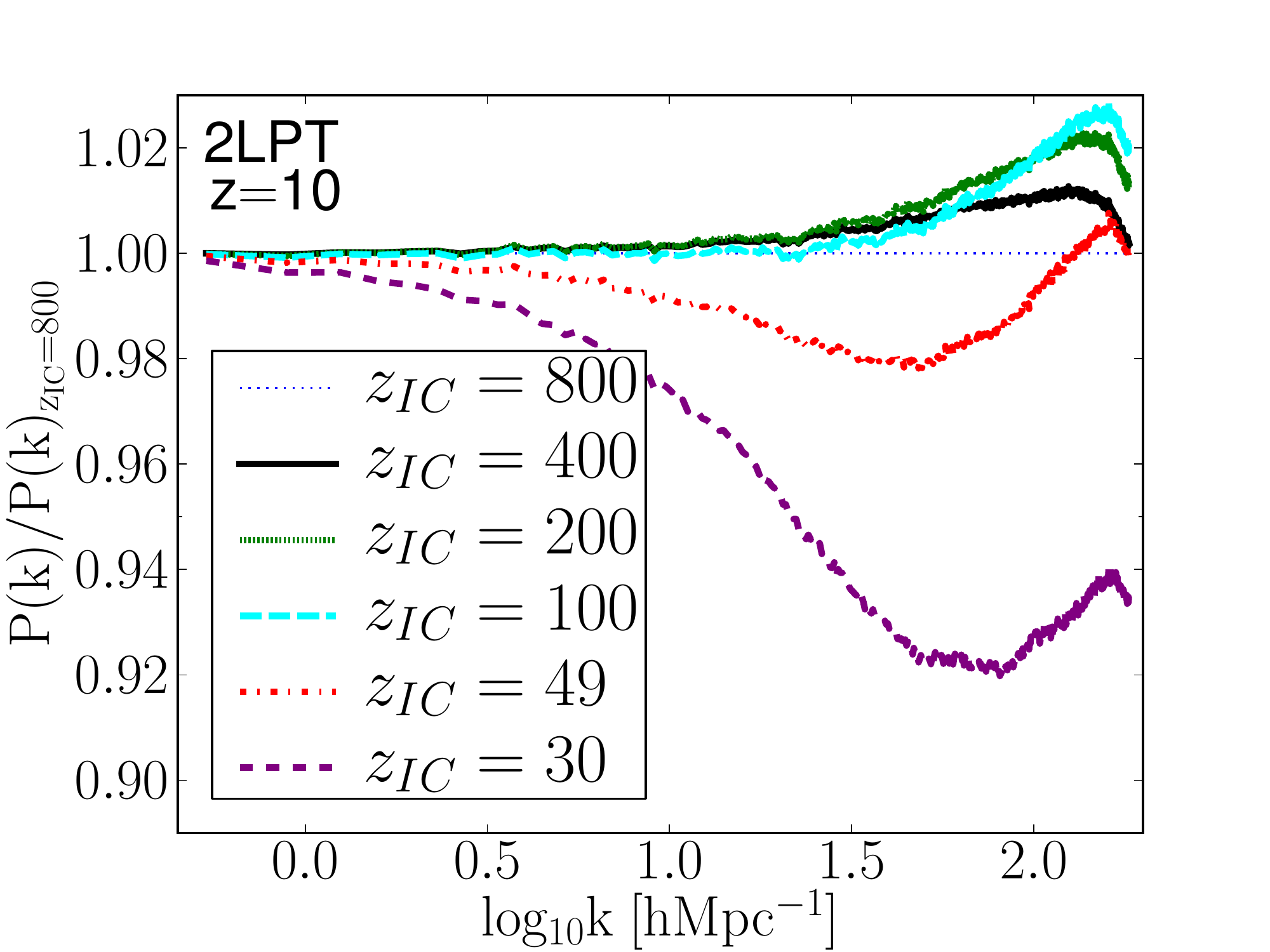}}
  \centering{
    \includegraphics[type=pdf,ext=.pdf,read=.pdf,height=0.38\textwidth,width=.5\textwidth]{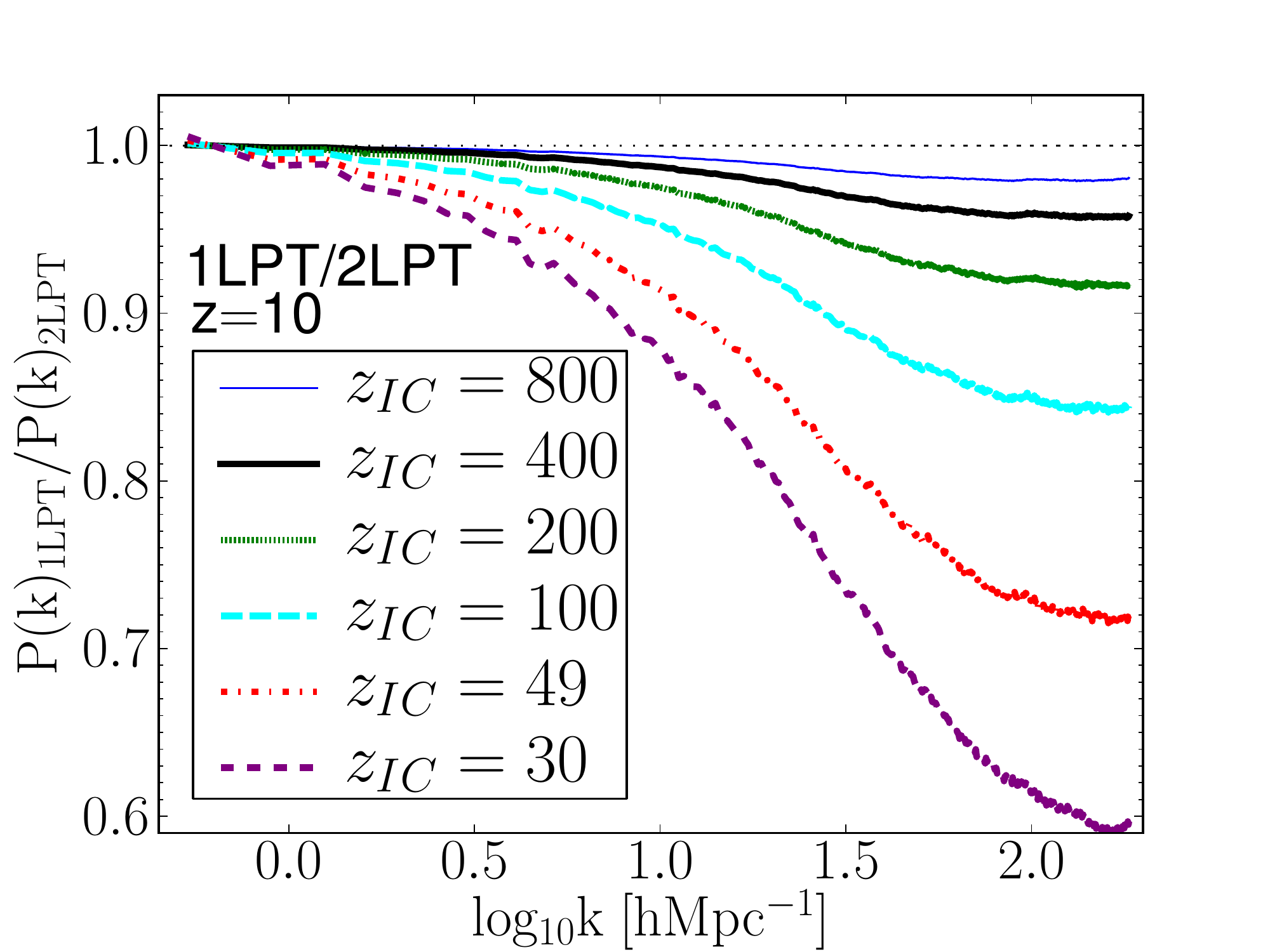}}
  \caption{Relative variation of the dark matter power spectra with
    the initial conditions for the $N$-body code {\tt PKDGRAV}. Top
    panel: variation of spectra with initial start redshift for 1LPT
    initial conditions. Middle panel: same as the top panel except for
    the case of the 2LPT initial conditions. Bottom panel: ratio of
    the power spectra from the 1LPT initial condition runs with
    respect to the 2LPT runs. All results shown are for the $L=17.625$
    box at $z=10$.  Percent level convergence is seen for 2LPT. }
  \label{fig-powz10pkd}
\end{figure}

%%%%%%%%%%%%%%%%%%%%%%%%%%%%%%%%%%%%%%%%%%%%%%%%%%%%%%%

\subsubsection{Variation with simulation parameters}

Figure~\ref{fig-powz10run} shows the dependence of the matter power
spectrum on the simulation run parameters: (top panel) tree-opening
angle $\Theta$; (middle panel) the force softening parameter
$\epsilon$; (bottom panel) time-step parameter $\eta$, for the
$N$-body simulation code {\tt PKDGRAV}.  These results show that the
estimated power spectra, on large scales\footnote{Note that owing to
  the fact that we are comparing results from the $L=17.5\Mpc$ at
  $z=10$ and $L=2048\Mpc$ boxes at $z=0$, we shall refer to
  wavenumbers in units of the fundamental frequency $k_{\rm
    fun}=2\pi/L$.} ($k/k_{\rm fun}<10$), are only weakly sensitive to
variations in the choice of $(\Theta, \epsilon,\eta)$. However, on
smaller scales, the power spectra show significant deviations. For the
force-softening tests, we find that for $\epsilon \le \lm/5$ the
spectra appear to be converged at the sub-percent level on large
scales, with a `bump' at $k/k_{\rm fun} \approx 100$ and a steep drop
at smaller scales.  This small-scale drop in power is consistent with
the puffing up of halo cores that appears to affect the mass function
when softening is large.  Similarly, for the case of the time-stepping
parameter $\eta$, we see that for $\eta=0.6$ there is a $\gtrsim1\%$
suppression of power for $k/k_{\rm fun}\gtrsim 100$. This can be
attributed to the large time-step not being able to follow the rapid
changes in the acceleration of particle orbits in the cores of
halos -- and hence the failure to capture the complex orbit of particles
in dense environments.  This discussion is limited to {\tt PKDGRAV} run parameters; a detailed study of the dependence of the power spectrum on {\tt Gadget-2} run parameters can be found in \citet{Smithetal2012}.

%%%%%%%%%%%%%%%%%%%%%%%%%%%%%%%%%%%%%%%%%%%%%%%%%%%%%%%

\subsubsection{Variation with initial conditions: small boxes}

Figure~\ref{fig-powz10pkd} shows the variation of the mass power
spectra with the choice of 2LPT or 1LPT initial conditions, and with
the adopted initial start redshift for the small box simulations at
$z=10$. The top panel of \Fig{fig-powz10pkd} shows the results for the
1LPT initial conditions. We observe that the power spectra are only
converged on the largest scales. On smaller scales, we find that the
power increases with increasing start redshift and that the results
are almost converged at the percent level only after $a_{\rm f}/a_{\rm
  i}\gtrsim 80$ expansions. 

The middle panel of \Fig{fig-powz10pkd} shows the results for the 2LPT
initial conditions. Here we find that simulations that were started
with $z_{i}\gtrsim100$ are converged for $k/k_{\rm fun}\lesssim 90$.
For simulations that possess lower start redshifts we find again a
suppression of power, although the effect is much reduced when
compared to the equivalent 1LPT start redshift. On smaller scales,
$k/k_{\rm fun}\gtrsim 100$, we find that the power is 1\% converged
for $100 \le z_{i} \le 400$ ($10 \lesssim a_{\rm f}/a_{\rm i} \lesssim
40$), while the $z=800$ start has up to 2\% less power.

The bottom panel of \Fig{fig-powz10pkd} presents the ratio of the
power spectra obtained from the 1LPT simulations with the 2LPT power
spectra, for various start redshifts. We see that the convergence of
the results from the 1LPT simulations with the 2LPT simulations is
very slow, and that percent level convergence is only obtained for 
$z_{\rm i}\gtrsim800$. 

Before continuing, we note that, whilst it appears that percent level
convergence in the power spectra may be achieved between 1LPT and 2LPT
for very high start redshifts, we have already shown in \S\ref{sec-ic}
that such high start redshifts are too early to produce an accurate
mass function, owing to numerical noise. We are therefore cautious
about such convergence.

%%%%%%%%%%%%%%%%%%%%%%%%%%%%%%%%%%%%%%%%%%%%%%%%%%%%%%%

\begin{figure}
  \centering{ 
   \includegraphics[type=pdf,ext=.pdf,read=.pdf,width=.52\textwidth]{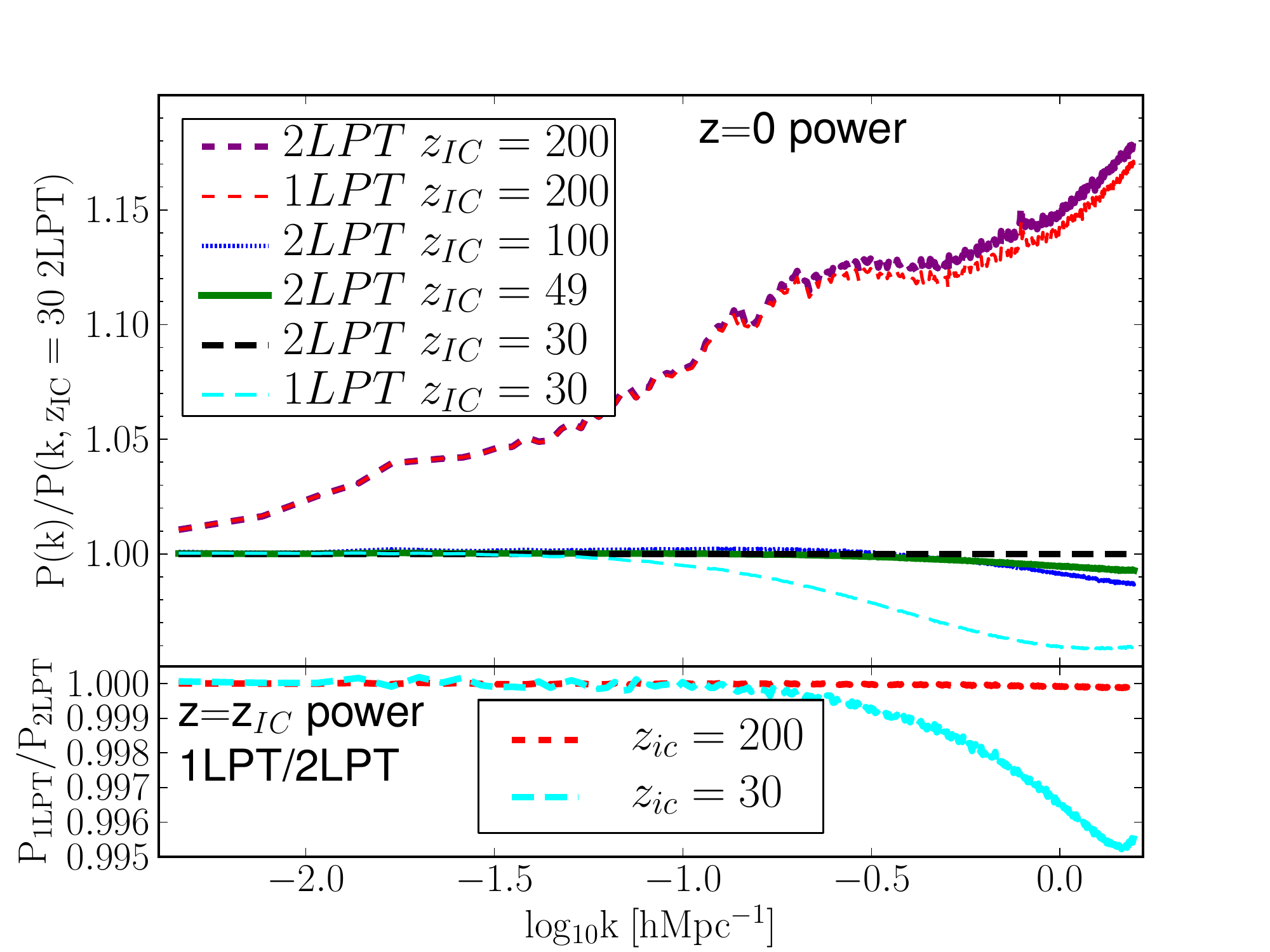}}
  \caption{ Top panel: Ratio of the evolved $z=0$ matter power spectra
    for runs with various initial start redshift $z_i$.  The evolved
    1LPT and 2LPT power spectra from the early $z_i=200$ start are
    very similar to each other, but they lie well above the converged
    results of lower redshift 2LPT starts.  Bottom panel: Initial
    ratio of the 1LPT to 2LPT matter power spectra at two selected
    initial start redshifts.  Differences in 1LPT and 2LPT power are
    very small ($<1\%$) at $z_i$, then grow much larger as the
    simulation evolves.  The simulations of this figure are large box
    $L=2048\Mpc$ simulations run with {\tt PKDGRAV}.}
  \label{fig-powz0pkd}
\end{figure}

%%%%%%%%%%%%%%%%%%%%%%%%%%%%%%%%%%%%%%%%%%%%%%%%%%%%%%%

\subsubsection{Variation with initial conditions: large boxes}

We now repeat the same set of tests as done for the previous
sub-section, only this time we now consider the $L=2048\Mpc$
simulation cubes at $z=0$.

Figure~\ref{fig-powz0pkd}, bottom panel, shows that the 1LPT and 2LPT
initial matter power spectra, measured at $z=z_{\rm i}$, are converged
at the 1\% level with respect to each other for the same start
redshift.  However, as indicated in the top panel of
\Fig{fig-powz0pkd}, the evolved spectra started with 1LPT are not
converged. On the other hand, the simulations started with the 2LPT
initial conditions, appear to be converged at the 1\% level for
$a_{\rm f}/a_{\rm i} \le 100$, except perhaps at the smallest scales,
even for start redshifts as low as $z_{\rm i}\sim30$. We note that the
evolved 1LPT versus 2LPT simulations, started with $z_{\rm i}=200$,
are significantly discrepant with respect to the other results. As for
the case of the mass function, we conjecture that this start redshift
is too early for the code {\tt PKDGRAV} to produce an accurate
integration of the equations of motion.  This reinforces our earlier
findings that 1LPT initial conditions are inadequate for accurate
simulations.

Several earlier studies have investigated the importance of 1LPT/2LPT
initial conditions on the matter power spectrum
\citep{Crocceetal2006,Heitmannetal2010}. Our findings are broadly
consistent with these studies. However, \citet{Heitmannetal2010}
advocated that 1LPT initial conditions started from $z_{\rm i}=200$
would lead to better than 1\% precision matter power spectra. Clearly
such a statement is code and run parameter dependent, and one should
be careful of increased numerical errors that may allow consistency
between 1LPT and 2LPT while still resulting in inaccurate power
spectra.

%%%%%%%%%%%%%%%%%%%%%%%%%%%%%%%%%%%%%%%%%%%%%%%%%%%%%%%

\begin{figure}
  \centering{
    \includegraphics[type=pdf,ext=.pdf,read=.pdf,width=.5\textwidth]{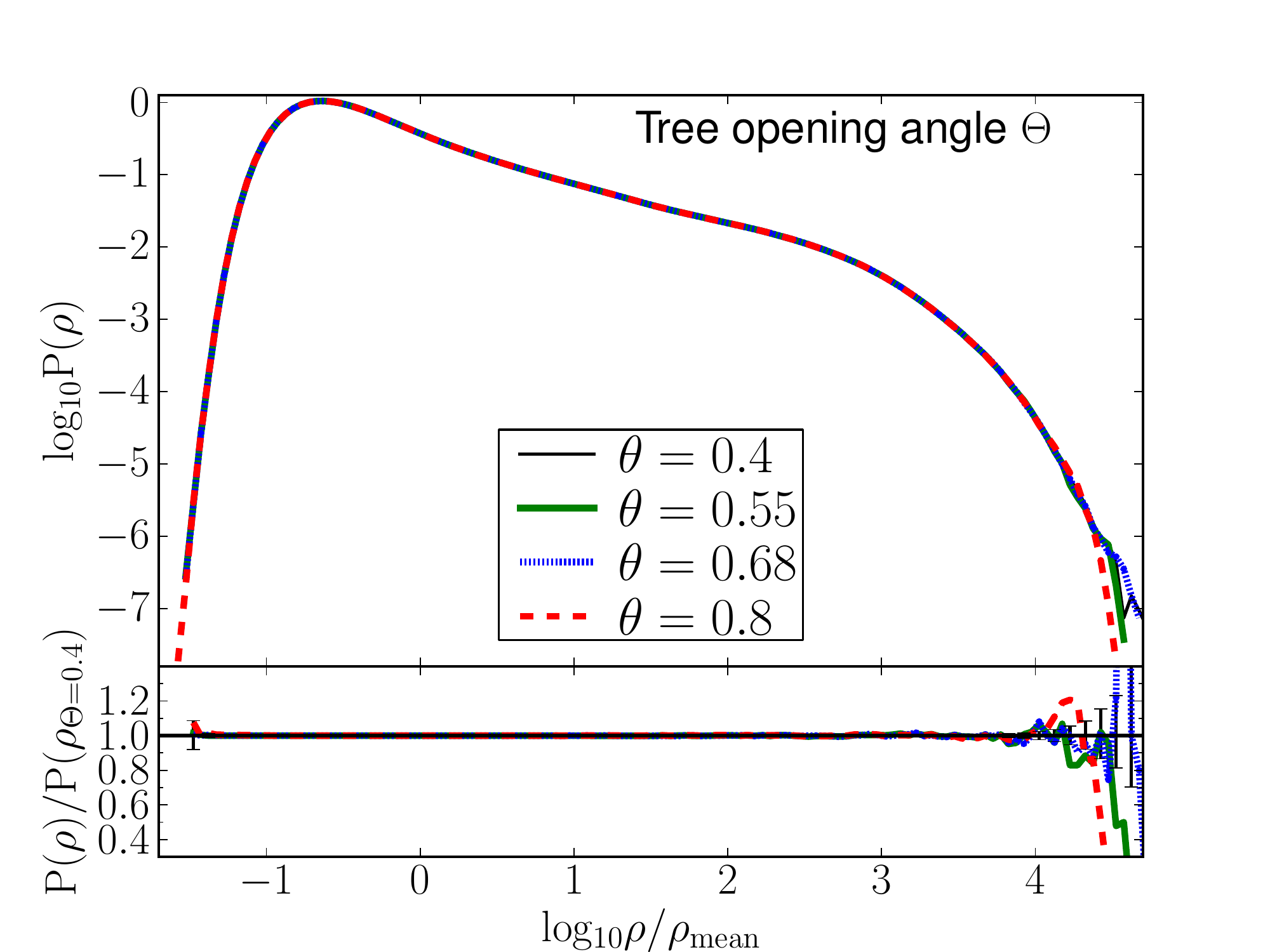}}
  \centering{
    \includegraphics[type=pdf,ext=.pdf,read=.pdf,width=.5\textwidth]{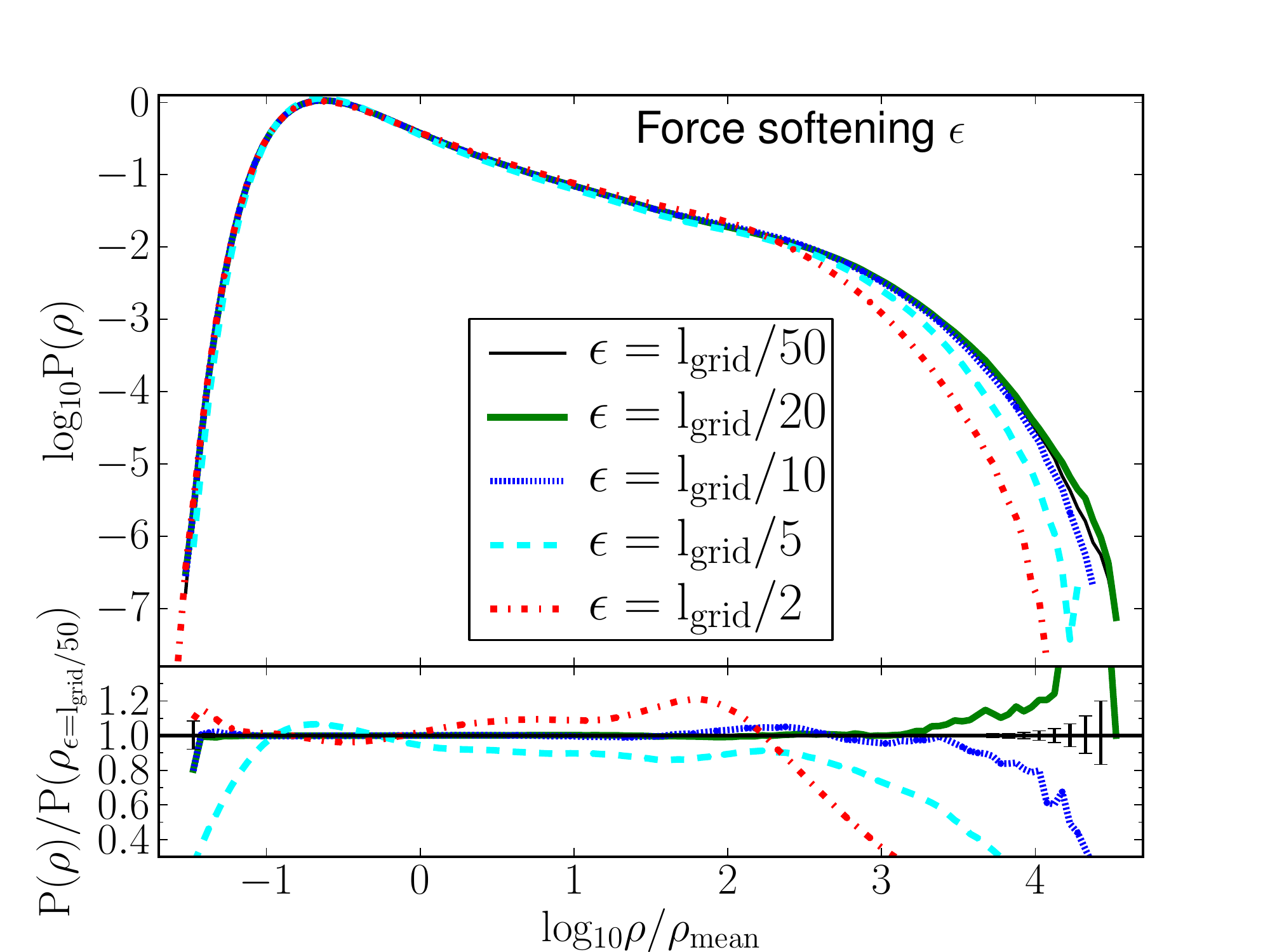}}
  \centering{
    \includegraphics[type=pdf,ext=.pdf,read=.pdf,width=.5\textwidth]{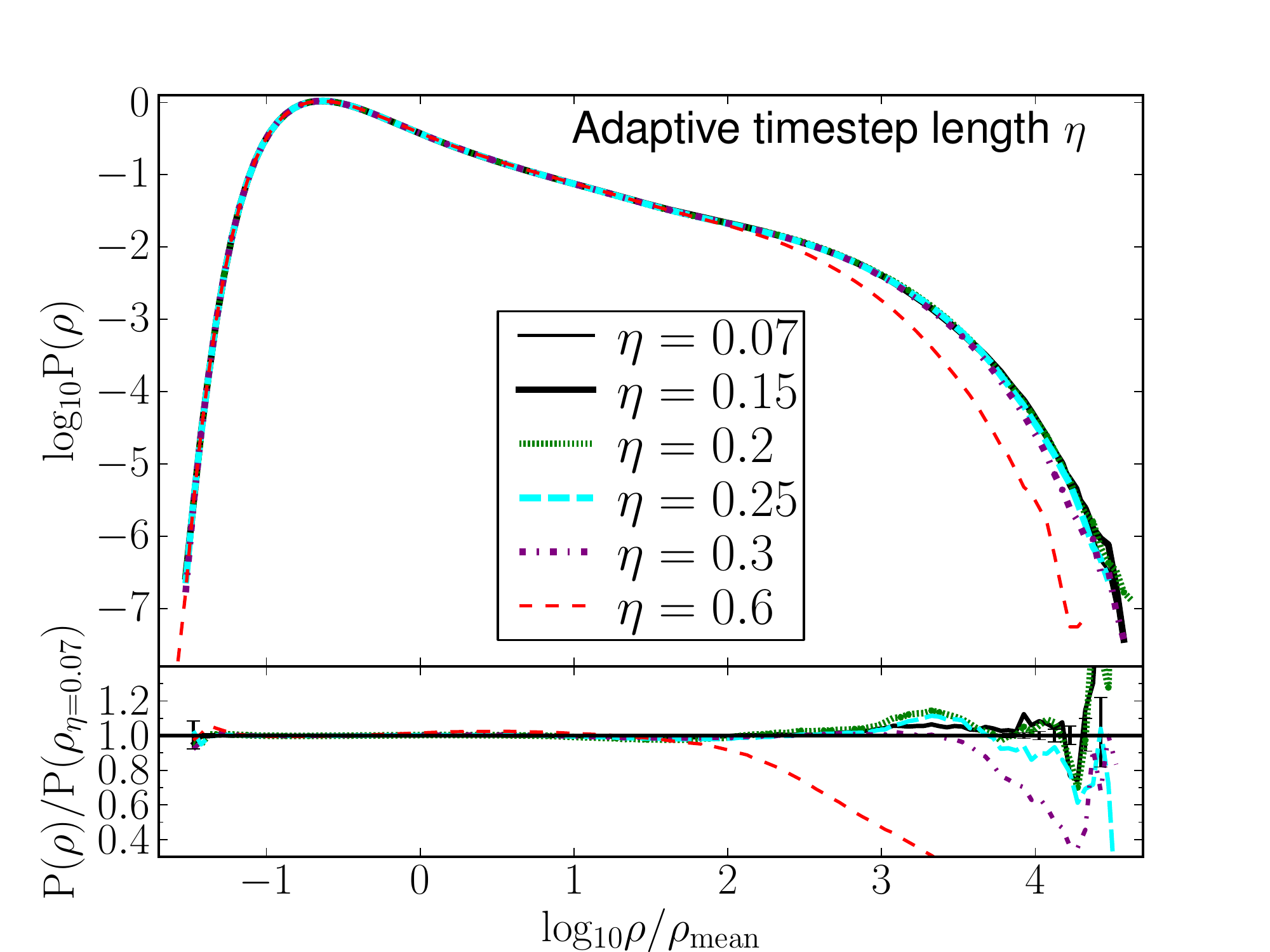}}
  \caption{Variation of the 1-point PDF of dark matter density
    fluctuations with the simulation parameters: Top panel, tree-opening angle $\Theta$; Middle panel, force softening $\epsilon$; and Bottom
    panel, time-stepping parameter $\eta$. All
    results here were obtained from the small $L=17.625\Mpc$ simulations at
    $z=10$ using the tree-code {\tt PKDGRAV}.  Sensitivity to numerical parameters is greatest at the highest densities.}
  \label{fig-rhorun}
\end{figure}

%%%%%%%%%%%%%%%%%%%%%%%%%%%%%%%%%%%%%%%%%%%%%%%%%%%%%%%

\begin{figure}
  \centering{
    \includegraphics[type=pdf,ext=.pdf,read=.pdf,width=.5\textwidth]{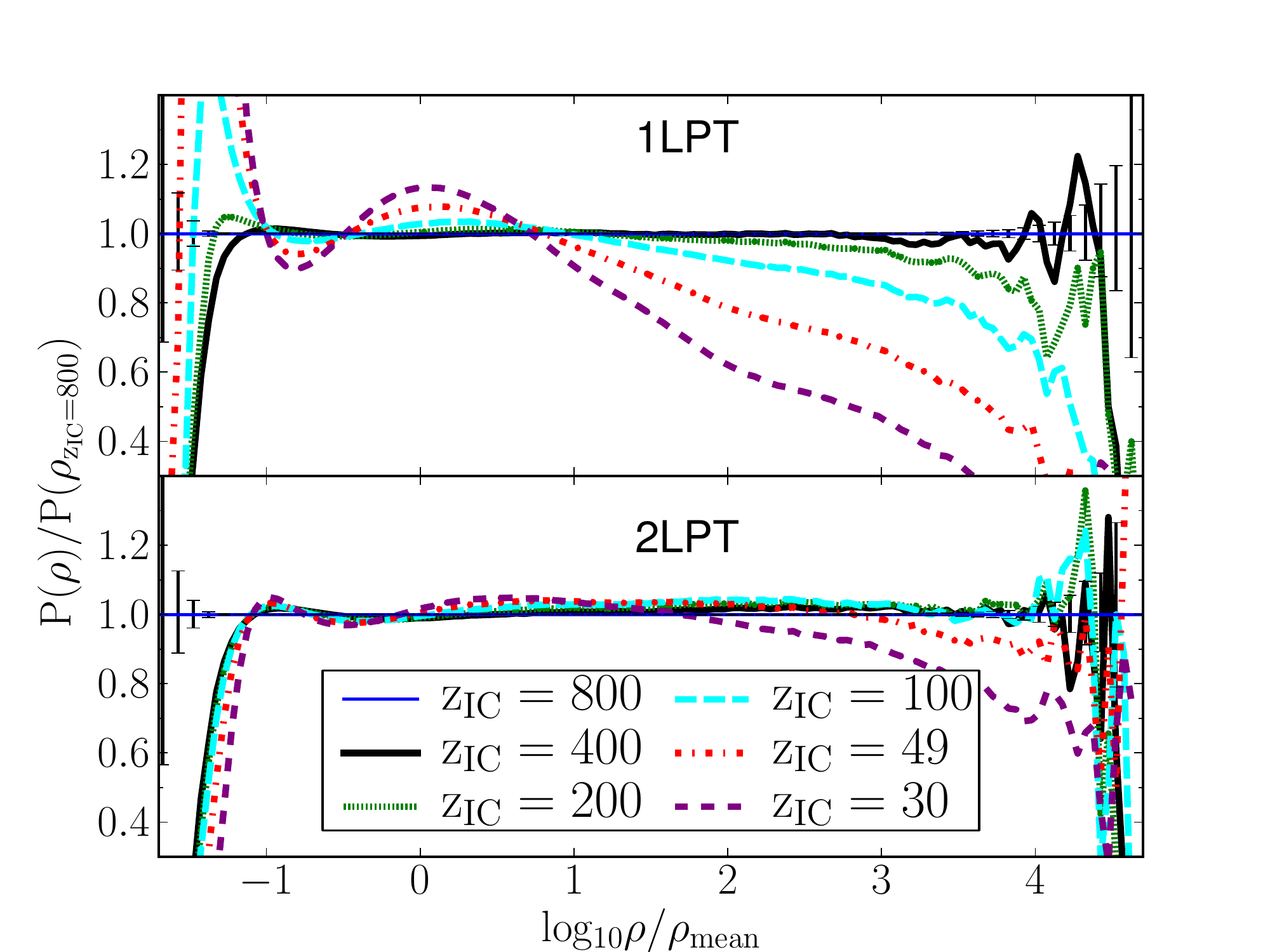}}
  \caption{Variation of the 1-point PDF of dark matter density
    fluctuations with respect to the choice of either 1LPT or 2LPT
    initial conditions for various start redshifts. Top panel: results
    for the simulations started with 1LPT initial conditions.  Bottom
    panel: shows the same but for the case of 2LPT initial
    conditions. These results were obtained from the small
    $L=17.625\Mpc$ simulations at $z=10$ using the tree-code {\tt
      PKDGRAV}.}
  \label{fig-rho1lpt2lpt}
\end{figure}

%%%%%%%%%%%%%%%%%%%%%%%%%%%%%%%%%%%%%%%%%%%%%%%%%%%%%%%

\subsection{1-point PDF of matter fluctuations}
\label{sec-rhoeps}

We now investigate the impact of simulation parameters and 1LPT versus
2LPT initial conditions on the 1-point PDF. At high densities, the
1-point PDF is a useful probe of the central regions of dark matter
halos, reflecting many properties of a ``stacked'' halo density
profile \citep[e.g.][]{ScherrerBertschinger1991}.  There are some
technical subtleties as to how one computes the 1-point PDF, since it
requires an estimator for the matter density at a given point. The
procedure of estimation in general requires one to smooth the particle
distribution and hence the results depend up on adopted smoothing
scale \citep[see for example][]{WattsTaylor2001}. Here we have chosen
to compute the 1-point PDF with a 64 particle nearest neighbor
kernel. This operates in a similar way to the SPH-kernel and
constitutes an adaptive smoothing scale.

Figure~\ref{fig-rhorun}, top panel, shows the variation of the 1-point
PDF with the tree-opening angle parameter $\Theta$.  We note that the
most significant changes are in the regions of highest density, though
sensitivity to $\Theta$, beyond the statistical fluctuations, is
relatively low.  Figure~\ref{fig-rhorun}, middle panel, shows the
variation of the 1-point PDF with the force softening $\epsilon$.
This clearly shows that the effect of too large force softening is to
damp the density distribution in the highest density inner regions of
dark matter halos.  This ``puffs up'' halos, which may explain the
increased abundances of lower mass halos with increased softening
length.  We also note that as $\epsilon=l/50$ the dense regions appear
to be again suppressed. This we attribute to violent two-body
encounters that can evaporate halo cores. Figure~\ref{fig-rhorun},
bottom panel, shows the variation of the 1-point PDF with the
time-stepping parameter $\eta$. We see that the results are well
converged provided $\eta\lesssim0.15$. The suppression of the high
density PDF for large $\eta$ reinforces our earlier speculation, that
if $\eta$ is too large, then the particle orbits in the cores of halos
can not be integrated sufficiently accurately, damping the
densities in the inner regions of dark matter halos.

Figure~\ref{fig-rho1lpt2lpt} presents the 1-point PDF for 1LPT and
2LPT initial conditions for various start redshifts.  As can be
clearly seen, the results for the 1LPT initial conditions converge
very slowly with start redshift. We also note that the both the high-
and low-density regions appear to be less dense for the simulations
that were started with low $z_i$. For the 2LPT simulations,
convergence is reached at much lower start redshifts, roughly $a_{\rm
  f}/a_{\rm i}=10$ expansion factors of the cube (i.e. around $z_{\rm
  i}\sim 100$).

We note all of the converged parameter values, are broadly consistent
with those that we identified for the mass function in \S\ref{sec-mf},
though variations in the PDF at the highest densities are generally
larger than 1\%.

%%%%%%%%%%%%%%%%%%%%%%%%%%%%%%%%%%%%%%%%%%%%%%%%%%%%%%%

\section{Discussion: remaining challenges for $<1\%$ accurate mass functions}
\label{sec-challenges}

In this section we discuss the remaining challenges that we will have
to face in order to approach better than 1\% accurate dark matter halo
mass functions.

%%%%%%%%%%%%%%%%%%%%%%%%%%%%%%%%%%%%%%%%%%%%%%%%%%%%%%%

\subsection{Mass resolution}
\label{sec-massres}

In the suite of tests above, we have seen indirect evidence that halos
with fewer than $N\sim1000$ particles, are unlikely to be useful for
deriving high accuracy estimates of the mass function.  This suggests
that there is a critical mass resolution, below which systematic
numerical errors are difficult to control.  Interestingly, this
resolution is somewhat worse than the $N\sim300$ particle resolution
limit expected for a pure particle-mesh (PM) code with mesh spacing
equal to the initial inter-particle separation
\citep{Lukicetal2007}. It is however still better than the more
conservative value of $N\sim2000$ particles proposed by
\citet{Bhattacharyaetal2011}. Additional support for our claim, comes
from the work of \citet{Trentietal2010}, who show through
mass-resolution tests, that on a halo-by-halo basis, halo masses with
$N\lesssim 1000$ particles are systematically too low.  Our
statistical limitations mean we can not rule out the possibility that
halos resolved with more $N\gtrsim3000$ particles might be accurate
over a larger range in $a_{\rm f}/a_{\rm i}$ than what we find for smaller
halos. In a subsequent paper, we will examine directly the mass
resolution convergence.

If the critical resolution limit of $N\sim1000$ particles for accurate
halo statistics is upheld, then this suggests that the tree code
technique may not have much advantage over the PM code technique in
recovering an accurate gravity-only mass function. Of course the
higher force resolution for tree codes enables better modelling of
higher density regions.

%%%%%%%%%%%%%%%%%%%%%%%%%%%%%%%%%%%%%%%%%%%%%%%%%%%%%%%

\begin{figure}
  \includegraphics[type=pdf,ext=.pdf,read=.pdf,width=.5\textwidth]{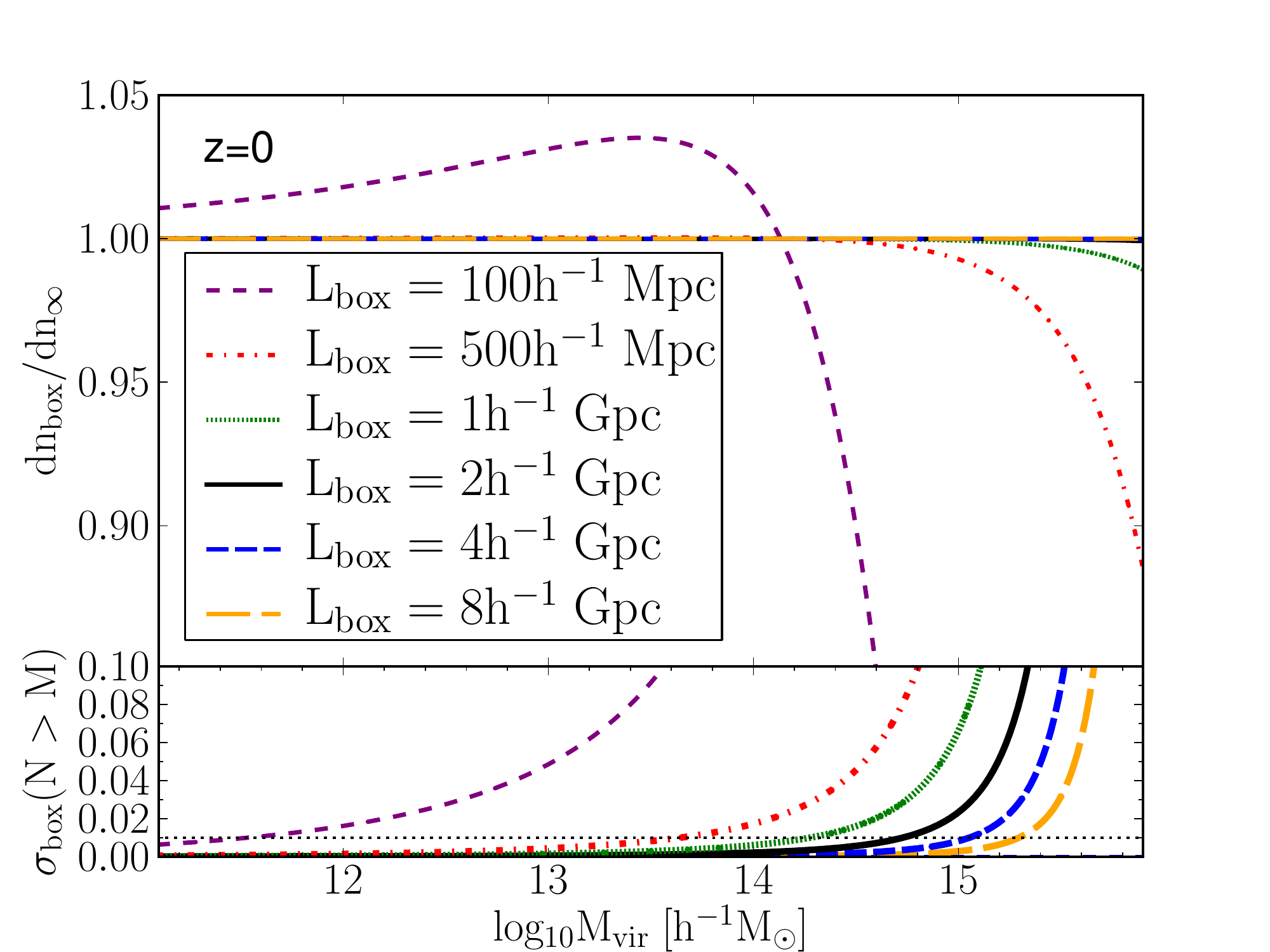}
  \caption{Estimates of the relative dependence of the mass function
    on simulation cube length compared to the case of an infinite
    cube. The dependence of the mass function on $L$ is calculated by
    assuming that the power on scales $k<k_{\rm fun}=2\pi/L$ is zero.
    The bottom panel is the expected Poisson error of the cumulative
    halo number count within a single simulation volume.  Predictions
    were obtained using the mass function of
    \citet{Bhattacharyaetal2011}.}
  \label{fig-boxsize}
\end{figure}

%%%%%%%%%%%%%%%%%%%%%%%%%%%%%%%%%%%%%%%%%%%%%%%%%%%%%%%

\subsection{Statistical precision}

The full-sky volume out to $z=2$ is $V_{\mu}\simeq 200\Gpccube$. This
sets the requirement on the minimum simulation volume needed to
replicate the survey volume accessible to future cluster surveys. This
would correspond to a single simulation cube with a side length of
roughly $L\gtrsim3.67\Gpc$, or if one wants to replicate the full
sky-light cone, then one would require a cube of length $L\sim8\Gpc$.
Obviously performing such a huge simulation with sufficient mass
resolution to obtain $N\sim1000$ particles per halo, for all halos
with masses $M\gtrsim 10^{13.5}\Msol$ would be very challenging
prospect. Such total volumes may be cheaply covered by combining the
results from many smaller volume simulations. However, individual
simulation boxes must be large enough to avoid systematic errors due
to mode-discreteness near the box-scale and due to the lack of super-box scale
power \citep[a ``DC-mode'' can help for multiple realization ensembles
][]{Sirko2005}.

An estimate of the minimum box size required to avoid suppressing
massive halo formation can be made by computing the effect of missing
power at wavelengths larger than the box on the (empirically fit)
analytic form of the mass function. Figure~\ref{fig-boxsize} (top
panel) demonstrates that a simulation box of roughly $L\sim2\Gpc$ box
should be able to capture the mass function at sub-percent level for
all halos with masses $M\lesssim 10^{16}\Msol$ at $z=0$. However, in a
single realization of such a volume, statistical accuracy will be much
lower; Poisson errors remain larger than $1\%$ well below
$10^{15}\Msol$.  One caveat is that this calculation underestimates
the finite volume effects due to the discreteness of modes near the
box scale \citep[see for example][]{Reedetal2007b,SmithMarian2011}.

%%%%%%%%%%%%%%%%%%%%%%%%%%%%%%%%%%%%%%%%%%%%%%%%%%%%%%%

\subsection{Verification of absolute accuracy}

The task of verifying that we have actually obtained the true answer
is somewhat circular, since it implies that we know already what the
true answer is. On our path toward the true answer, we should
consider the possibility that our simulations suffer from some level
of false convergence.  One obvious approach to addressing this issue
will be to verify that independent simulation codes give the same
results at the desired level of accuracy. However, this does not take
into account the pernicious systematic errors, such as the false
convergence with redshift that we observed for the 1LPT simulations.
In the future, a more complete approach would provide us with a
theoretical framework for objectively quantifying `accuracy' and
enable us to identify directions in parameter space that would allow
us to approach our desired goal.

There are a number of systematic errors that one will need to
characterise in detail. One in particular is that associated with the
coarse graining of phase space.  Spurious perturbations related to mass
discreteness may lead to the collapse of small structures around
lattice points \citep{Melott1990,JoyceMarcos2007,Joyceetal2009}. This
effect is well-known for Warm Dark Matter simulations
\citep{WangWhite2007,Schneideretal2012}, and it therefore must also be
present in CDM simulations.  
Other effects relating to mass discreteness in particle codes have been 
discussed by a number of authors \citep[\eg][]{Splinter1998,JoyceMarcos2007b,Romeoetal2008}.

We have attempted to steer away from making direct statements regarding
whether particular papers had errors and by how much because
quantifying the accuracy of the works of other authors would require
repeating their simulations with exactly the same codes and run
parameters.  A number of widely used fits in the literature used 1LPT
initial conditions, many of them with a lower start redshift than
expected to be required for good agreement between 1LPT and 2LPT ICs.
In those cases, a systematic under-abundance of the halo mass function,
especially at high masses or high redshifts is implied.

An informative comparison can be made from the fact that the fitting
formulae of multiple authors are in reasonable agreement with each
other.  For example, there is agreement at the $2-3\%$ level between
the FoF mass function fits of \citet[][using 2LPT
  ICs]{Crocceetal2010}, \citet[][using a mix of 2LPT and 1LPT with a
  high start redshift]{Bhattacharyaetal2011} and \citet[][using
  2LPT]{Anguloetal2012}, over a wide mass range.  Also, there is
$\simeq5\%$ agreement between the \citet[][1LPT ICs]{Tinkeretal2008}
and the recent \citet[][1LPT ICs with a higher start
  redshift]{Watsonetal2012} spherical over-density mass function fits.
There is a caveat that such comparisons are only useful to the extent
that the different studies do not suffer from the same systematic
errors.

If we consider the widely-used \citet{Tinkeretal2008} mass function,
the authors stated statistical accuracy of $5\%$ is comparable to the
systematic error, estimate from Eqn. \ref{eqn-unifit}, that we would
expect from their results due to their use of 1LPT initial conditions
with $a_{\rm f}/a_{\rm i} \sim 50$, although this error would approach
$10\%$ at the highest masses, while being smaller at lower masses
($\simlt 10^{15} \Msol$).  They point out some dependence on start
redshift between some of their simulations, and exclude from their fit
those with the lowest start redshifts due to a systematic
under-abundance of halos.  We can thereby deduce that the systematic
errors in that study due to initial conditions are generally within
their quoted statistical accuracy.

%%%%%%%%%%%%%%%%%%%%%%%%%%%%%%%%%%%%%%%%%%%%%%%%%%%%%%%

\subsection{Narrow scale factor range for accuracy}

In \S\ref{sec-ic}, we showed that significant errors are introduced in
the halo mass function when the total expansion of the box lies
outside the limits $(10\lesssim a_{\rm f}/a_{\rm i}\lesssim50)$.  
Of course, one would expect that the epoch of a particular halo's formation is more important than final simulation output for assessing whether that halo, and by extension, the mass function, has been modelled accurately.
A brief supporting
argument is that any errors introduced to early structures are
unlikely to ``evolve away'', though suppression of the number of early
forming halos may become less noticeable, at fixed mass, after more
halos have formed at lower redshifts.  The implication is that halos
in any particular cosmological simulation can be modeled accurately
only for a range in formation redshifts of ${\Delta_{1+z} \sim 5}$.
It is possible that this restriction may become less severe as the
mass resolution of the simulation is increased. The logic being that
more cosmological power at small scales may enable earlier start
redshifts to be simulated without leading to an increase in the
amount of spurious structure.  And, as shown in \S~\ref{sec-ic}, the mass 
function sensitivity to start redshift is smaller for halos resolved with
more particles.  
The upper limit of the allowed start redshift range may be code
dependent.  For example, as we discussed in \S\ref{sec-largeboxes}, a
particle-mesh technique may allow higher start redshifts, but
typically comes at the cost of worse force resolution.

The narrow range in implied $a_{\rm f}/a_{\rm i}$ presents a challenge
for simulations with very large dynamic range, wherein it would be
difficult to model accurately the mass function of massive cluster
halos forming at $z \sim 0$, while also capturing accurate evolution
of the early generations of dwarf galaxy halos forming already at $z
\sim 10$. Even though the fraction of mass assembled into galactic
halos at such early times is small, early galaxy formation occurs
preferentially in Lagrangian regions where clusters will later form.
These early-forming galactic halos should be modeled accurately
because their feedback processes could have have significant effects
on the eventual baryon and total cluster mass content through energy
injection and preheating, which could begin very early
\citep[\eg][]{BensonMadau2003}.

%%%%%%%%%%%%%%%%%%%%%%%%%%%%%%%%%%%%%%%%%%%%%%%%%%%%%%%

\subsection{Impact of baryons}

We have purposely ignored the important effects of baryons on the halo mass
function.  Recent hydrodynamic simulations have shown the range of
plausible baryon effects on total cluster halo masses to be up to
$\simeq 15\%$ within the radius enclosing $500\times$ the critical 
density of the universe
% \rho_{\rm c}$
\citep{Staneketal2009}.  Even for ``adiabatic simulations'' wherein
gas cooling, star formation and feedback are ignored, baryons may have
up to $\sim 7\%$ effects on the halo mass function
\citep{Cuietal2012}.  

Baryon influences thus present a serious challenge for percent level
accuracy in the mass function required for planned dark energy
missions.  However, accurate gravity-only simulations, as we have
explored in this work, are a pre-requisite for future simulations with
more complete baryon physics aiming to obtain percent level accuracy
in the mass function or other properties.  To save computational cost,
cosmological constraints might rely on a combination of baryon and
gravity-only simulations.  For example, hydrodynamic simulations of
limited cosmological volumes could be used to calibrate the effects of
baryons on halos as well as to derive relations between halos and
observable properties.  Then, large volume gravity-only simulations
might be utilized to map the dependence of halo numbers and other
properties on cosmology.

%%%%%%%%%%%%%%%%%%%%%%%%%%%%%%%%%%%%%%%%%%%%%%%%%%%%%%%

\subsection{Calibrating mass--observable relationship}

A further difficulty in using the halo mass function for cosmology is
that any practical halo definition one might use, such as through the
FoF or SO algorithms, typically have no direct observable counterpart.
For example, the emission from X-ray clusters is determined by the (square of)
the gas density distribution.  One
might argue that this leads to more spherical clusters the
abundance of which may be better matched to SO halos in simulations,
rather than to FoF halos (see discussion in \citealt{KravtsovBorgani2012}). 
Though weak lensing may help to calibrate
halo masses \citep{Marianetal2009,BeckerKravtsov2011,Mandelbaumetal2010}, optical, x-ray,
and Sunyaev Zel'dovich cluster masses have large scatter and
systematic uncertainties \citep{Anguloetal2012}. 

Mock observations from simulations may thus be a superior means of
obtaining a more accurate mass function in the observable
plane, especially once baryon properties are better modeled.
Ultimately, an observationally useful and accurate (cluster) mass
function will involve modeling baryon physics and observable
properties (or mock observations) and represents a formidable
challenge for cosmology -- a challenge that must be solved in order to
fully exploit future and even current cluster surveys of the Universe.

%%%%%%%%%%%%%%%%%%%%%%%%%%%%%%%%%%%%%%%%%%%%%%%%%%%%%%%

\begin{table*}
\begin{center}
\begin{tabular}{ccccccccc}
\hline
           & $\Theta$ & ${\rm \epsilon}$ & $\eta$ & $\eta_{\rm G}$ & $a_{\rm f}/a_{\rm i}$ & ${\rm N}$ per halo & $L\,[\Gpc]$ & $V_{\mu}(z<2)\,[\Gpccube]$ \\

\hline
min. value & --       & $\lm/50$         &   --   &         --    &       10           &        1000        &     2.0     &    $\sim$200  \\  
max. value & 0.7      & $\lm/20$         & 0.15-0.20  &        0.02   &       40-50           &         --         &     --      &     --    \\ 
\hline
\end{tabular} 
\caption{{\it Approximate} run and initial condition parameters that
  permit percent-level simulation convergence in the halo mass
  function extracted from gravity-only simulations.  2LPT initial
  conditions are required. Run parameters $\Theta$, $\epsilon$, and
  $\eta$ are the tree-opening angle, force softening, time-stepping
  parameter. $\eta_{\rm G}$ denotes the {\tt Gadget-2} time-step
  parameter ${\tt ErrTolIntAccuracy} =\eta^2/2$.  $a_{\rm f}/a_{\rm
    i}$ is the ratio of initial to final scale factor at which halo
  properties are to be considered.  $L$ is an estimate of the
  minimum box length needed to avoid systematic errors in the mass
  function while $V_{\mu}$ is the comoving volume of the universe
  accessible to future cluster surveys.}
\label{table-recipe}
\end{center}  
\end{table*}

%%%%%%%%%%%%%%%%%%%%%%%%%%%%%%%%%%%%%%%%%%%%%%%%%%%%%%%

\newpage
%\clearpage 
\section{Conclusions and guidelines for accurate simulations}
\label{sec-guidelines}

In this paper we have explored the dependence of the mass function of
dark matter halos on simulation run parameters and initial conditions.
Our aim has been to perform convergence tests that will illuminate the
path to obtaining percent level accuracy in this statistic. This will
be a requirement for future cluster surveys of the Universe that aim 
to help constrain the nature of dark energy or dark gravity. 

In \S\ref{sec-techniques} we gave a brief overview of the simulation
method, paying special attention to how one sets up initial
conditions, either using Zel'Dovich approximation (1LPT) or 2LPT. We
described the simulation codes that we have employed {\tt PKDGRAV} and
{\tt Gadget-2}, with the former being the main code used throughout
this study.

In \S\ref{sec-simulations} we described the large suite of $N$-body
simulations that we have performed to study these problems. All
simulations were run at $N=1024^3$ and we covered two regimes: high
redshift ($z=10$), small scale ($L=17.625\Mpc$) and low redshift
($z=0$) large scale ($L=2\Gpc$).

In \S\ref{sec-mf} we explored the dependence of the mass functions on
the simulation parameters and found the following: the resultant mass
functions were rather insensitive to the choice of the tree-opening
angle, provided $\Theta<0.7$; the results for halos resolved with
fewer than $N\sim1000$ particles were sensitive to the choice of force
softening, with larger values tending to increase the abundance of
halos in this regime; results were fairly insensitive to the size of
the adaptive time-step parameter and that 1\% converged results could
be achieved for $\eta\lesssim0.15$. We also demonstrated that the use
of anti-aliasing filters, such as the Hann filter, to set up initial
conditions, can lead to $\sim30\%$ suppression in the abundance of
halos resolved with $N\lesssim1000$ particles. We do not advocate the
use of the Hann filter, since {\em there is no aliasing} in the initial
conditions to correct.

In \S\ref{sec-ic} we performed a detailed study of the impact of the
choice of initial conditions on the mass function. We found that the
results from simulations that are initialized with 1LPT converge very
slowly as the start redshift is increased. The effect of too low a
start redshift being the suppression of the formation of high mass
halos. Furthermore, for the large box simulations, we also found
simulations started at very high redshifts $z_{\rm i}\gtrsim200$ would
fail to correctly follow the build up of structure due to the relative
increase in numerical noise. Furthermore, 1LPT initial conditions
exhibit ``false convergence'' with increasing start redshift.
Simulations starting from 2LPT initial conditions proved to have very
good convergence properties and for simulations that underwent 10-50
expansion factors, yielded percent level convergence in the halo mass
function at the $1024^3$ resolution of our tests. We made a direct
comparison of these results obtained from integration of the initial
conditions with the tree-code {\tt ~PKDGRAV} with results from the
Tree-PM code {\tt ~Gadget-2}, and found almost identical
behaviour. However, a detailed comparison of the mass functions from
the two codes revealed that {\tt Gadget-2} produced a $\sim10\%$
increase in the mass function for halos resolved with $N\lesssim10^2$
particles. These results extend and support the earlier findings of
\citep{Crocceetal2006}.

In \S\ref{sec-conother} we explored the convergence properties of two
other statistics of the density field, namely the matter power
spectrum and the 1-point probability density function (PDF) of matter
fluctuations. We found that the simulation parameters that produced
$\lesssim1\%$ convergence in the mass function would also lead to good
convergence behaviour in these statistics. In addition, too high a
start redshift for either 1LPT or 2LPT initial conditions would lead
to systematic errors. On the other hand, the results from the
simulations run with 2LPT initial conditions demonstrated excellent
convergence behaviour.

In summary, Table~\ref{table-recipe} presents a general recipe for the
parameters needed for percent accuracy of the mass function within a
gravity-only simulation using a tree code. Except for the tree
opening-angle $\Theta$, which has some dependence on the specific tree
used, all the other run parameters can be applicable to other tree
codes. This list shows required values but is not complete.  In future
work, one would expect this table to be extended to include the
following: if PM forces are used for large scale force computation,
then parameters controlling their accuracy, such as the size of the PM
grid should be included; multipole expansions are used to compute the
tree forces, and different codes use different orders: which order is
sufficiently accurate for our purposes? Also, there should be some
entry associated with the parameters that control the halo finder
(halo definition).

Ultimately, inferring cosmological parameters from the cluster mass
function will require a number of other issues to be solved relating
to baryons and observable properties.  Among the difficulties that
baryons pose is the gravitational coupling of baryon processes to dark
matter \citep{SomogyiSmith2010,vanDaalenetal2011}.  Inferring
observable properties from the simulations for comparison via
mass-observable relations or by direct mock catalogs is a further
complexity.  Thus, percent level accuracy in numerical simulations
represents a formidable challenge, but one that we must meet if future
surveys of the Universe are to live up to their potential.

%%%%%%%%%%%%%%%%%%%%%%%%%%%%%%%%%%%%%%%%%%%%%%%%%%%%%%%
%%%%%%%%%%%%%%%%%%%%%%%%%%%%%%%%%%%%%%%%%%%%%%%%%%%%%%%

\section{acknowledgments}
We thank the anonymous referee for helpful suggestions that have improved
this work.
We thank Martin Crocce, Cameron McBride, Roman Scoccimarro, and Romain
Teyssier for helpful discussions.  We also thank Salman Habib, Katrin
Heitmann, and Zarijah Luki{\'c} for early discussions that influenced
this work.  We thank Roman Scoccimarro for making {\tt 2LPT} public;
Volker Springel for making {\tt GADGET-2} public, and for providing
his {\tt B-FoF} halo finder; and Ed Bertschinger for making {\tt
  GRAFIC-2} public. The simulations were performed on Rosa at the Swiss
National Supercomputing Center (CSCS), and the zbox3 and Schr\"odinger
supercomputers at the University of Zurich. RES acknowledges support
from a Marie Curie Reintegration Grant and the Alexander von Humboldt
Foundation.

%%%%%%%%%%%%%%%%%%%%%%%%%%%%%%%%%%%%%%%%%%%%%%%%%%%%%%%

% BIBILIOGRAPHY

\bibliographystyle{hapj}
\bibliography{refsDarren}

%%%%%%%%%%%%%%%%%%%%%%%%%%%%%%%%%%%%%%%%%%%%%%%%%%%%%%%

\label{lastpage}
\end{document}

%% file: defsDarren.tex
\newcommand{\etal}{{et al.}~}

\newcommand{\eg}{{e.g.~}}
\newcommand{\ie}{{i.e.~}}

\def \ltsima{$\; \buildrel < \over \sim \;$}
\def \simlt{\lower.5ex\hbox{\ltsima}}            % < over ~
\def \gtsima{$\; \buildrel > \over \sim \;$}
\def \gtsima{\mbox{$\; \buildrel > \over \sim \;$}}
\def \simgt{\lower.5ex\hbox{\gtsima}}            % > over ~

\newcommand{\be}{\begin{equation}}
\newcommand{\ee}{\end{equation}}
\newcommand{\ba}{\begin{eqnarray}}
\newcommand{\ea}{\end{eqnarray}}

\newcommand\Eqn[1]     {Eq.\,(\ref{#1})}

\newcommand\Fig[1]     {Fig.\,{\ref{#1}}}

\newcommand\nn         {\nonumber}

\def\lm{{l_{\rm m}}}

\def\pp1{{\prime}}
\def\pp2{{\prime\prime}}

\def\2D{{\rm 2D}}

\def\bx{{\bf x}}

\def\bk{{\bf k}}
\def\bq{{\bf q}}

\def\1Loop{{\rm 1Loop}}

\def\rhob{\bar{\rho}}

\def\Msol{h^{-1}M_{\odot}}

\def\Mpc{\, h^{-1}{\rm Mpc}}

\def\Gpc{\, h^{-1}{\rm Gpc}}
\def\Gpccube{\, h^{-3} \, {\rm Gpc}^3}

\def\dx{{\rm d}^3{\!\bf x}}

\def\la{\mathrel{\mathpalette\fun <}}
\def\ga{\mathrel{\mathpalette\fun >}}
\def\fun#1#2{\lower3.6pt\vbox{\baselineskip0pt\lineskip.9pt
        \ialign{$\mathsurround=0pt#1\hfill##\hfil$\crcr#2\crcr\sim\crcr}}}